\documentclass[aps,prc,notitlepage,floatfix,nofootinbib,amsmath,amssymb,amsfonts, twocolumn]{revtex4-2}

\usepackage[utf8]{inputenc}
\usepackage[T1]{fontenc}
\usepackage{lmodern}
  
\usepackage{mathrsfs}
\usepackage{graphicx}
\usepackage{epsfig}     
\usepackage{bm}
\usepackage{color}
\usepackage{float}
\usepackage{dcolumn}
\usepackage{multirow}
\usepackage{xcolor}
\usepackage{hyperref}
\hypersetup{colorlinks,
            linkcolor={red!50!black},
            citecolor={blue!50!black},
            urlcolor={blue!80!black}
}
\usepackage{color}
\usepackage{braket}
\usepackage{amsmath}
\usepackage{enumitem}

\begin{document}

\title{How Bayesian methods can improve \textit{R}-matrix analyses of data: the example of the \textit{dt} Reaction}

\author{Daniel Odell}
	\affiliation{Institute of Nuclear and Particle Physics and Department of Physics and Astronomy, Ohio University, Athens,
		Ohio 45701, USA}

\author{Carl~R. Brune}
	\affiliation{Institute of Nuclear and Particle Physics and Department of Physics and Astronomy, Ohio University, Athens,
		Ohio 45701, USA}

\author{Daniel~R. Phillips}
	\affiliation{Institute of Nuclear and Particle Physics and Department of Physics and Astronomy, Ohio University, Athens,
		Ohio 45701, USA}

\begin{abstract}
The $^3{\rm H}(d,n)^4{\rm He}$ reaction is of significant interest in
nuclear astrophysics and nuclear applications.
It is an important, early step in big-bang nucleosynthesis and 
a key process in nuclear fusion reactors.
We use one- and two-level $R$-matrix approximations to analyze
data on the cross section for this reaction at center-of-mass energies below 215 keV.
We critically examine the data sets using a Bayesian statistical model that allows for both common-mode and additional point-to-point uncertainties. We use Markov Chain Monte Carlo  sampling to evaluate this $R$-matrix-plus-statistical model and find two-level $R$-matrix results that are stable with respect to variations in the channel radii. The $S$ factor at 40 keV evaluates to $25.36(19)$ MeV b (68\% credibility interval). We discuss our Bayesian analysis in detail and provide guidance for future applications of Bayesian methods to $R$-matrix analyses. 
We also discuss possible paths to further reduction of the $S$-factor uncertainty.
\end{abstract}
 
\maketitle

\section{Introduction}

The $^3{\rm H}(d,n)^4{\rm He}$ reaction has been measured a number of times at low energies. Data sets with small statistical error bars exist that, taken together, cover the energy range 5-250~keV\footnote{All energies in this paper refer to the c.m. system, unless otherwise indicated.} including the broad resonance at $E\approx 90$~keV~\cite{Con52,Arn53,Kobzev1966,Jar84,Bro87}.

Accurate knowledge of this resonance is important because the large low-energy cross section associated with it, together with the $Q$~value of the reaction of 17.6 MeV, mean the $dt$ reaction is an efficient approach for energy generation from nuclear fusion. For commercial fusion reactor purposes this knowledge is needed at energies below 100 keV. The $^3{\rm H}(d,n)^4{\rm He}$ reaction also plays a role in Big Bang Nucleosynthesis. In that case the relevant energy range is a bit larger~\cite{Nol00}, extending to the upper end of the domain covered by the experiments listed above. 

A recent {\it ab initio} calculation provides a good description of the data in the energy region of interest~\cite{Hupin:2018biv}. But even this calculation requires phenomenological adjustment if it is to provide an accurate description of the low-energy $dt$ data. 
The presence of a single, broad resonance in this kinematic domain has made $R$-matrix methods the tool of choice for evaluation and extrapolation of the $^3{\rm H}(d,n)^4{\rm He}$ cross section---although Effective Field Theory also provides a simple parameterization that can accommodate these data~\cite{Brown:2013zla}.
The most sophisticated $R$-matrix analyses of this reaction are by \textcite{Hale:1987} and \textcite{Bosch:1992,Bosch:1993}. They yield a value for the $S$ factor of the reaction at the canonical $E=40$ keV evaluation point of $25.87 \pm 0.49$~MeV b~\cite{Bosch:1993}. 
More recently \textcite{deSouza:2019pmr} employed a Bayesian methodology and a one-level $R$-matrix approximation to study  the $dt$ reaction in the 0--250 keV energy range. Ref.~\cite{deSouza:2019pmr} used the $R$-matrix parameterization of Lane and Thomas~\cite{LaneThomas1958} and computed the posterior probability density (pdf) of the $R$-matrix parameters in a seven-dimensional parameter space. This allowed de Souza {\em et al.} to straightforwardly propagate the uncertainties in parameters to final results for the $S$~factor. It also made it straightforward for them to introduce what we shall refer to as a ``statistical model'' to go along with their one-level $R$-matrix physics model. That statistical model accounted for imperfections in the experiment by introducing ``nuisance'' parameters associated with the normalization error in a particular data set, systematic shifts in the actual energies compared to those quoted, additional point-to-point errors beyond those quoted in the original papers, etc. In a Bayesian framework the (several) additional parameters introduced to account for the possibility that such effects are present can be marginalized over, meaning that the corresponding uncertainty is included in the final error bar for $S(40~{\rm keV})$. Of particular note is de Souza {\em et al.}'s finding that the data sets referenced in the first paragraph should have their point-to-point $S$-factor errors enhanced by as much as 0.5 MeV b if they are to be statistically consistent with one another. 

In this work we adopt a similar strategy to de Souza {\em et al.}, using Markov Chain Monte Carlo (MCMC) sampling to explore the posterior pdf of the combination of $R$-matrix and statistical parameters. We improve upon their study in several ways. First, we point out that two of the $R$-matrix parameters de Souza {\em et al.} sampled are degenerate with regard to their impact on the reaction cross section. We therefore eliminate one. Second, we show that including the channel radii as parameters results in extended non-linear correlations in parameter space. This makes MCMC sampling slow to converge. Our sampling uncovers these structures, which are not present in the posteriors of de Souza {\em et al.}~\cite{deSouza:2019pmr}. 
But the channel radius is not a parameter: it is instead a regulator that separates internal dynamics from the asymptotic wave functions. 
An $R$-matrix with an infinite number of levels should produce results for observables that are independent of the channel radius. 
The channel radius should thus not be thought of as a parameter to be optimized or sampled; instead observables should be largely independent of it.
Third, and relatedly, we investigate the limitations of de Souza {\em et al.}'s assumption that this reaction can be described by a single $R$-matrix level. We include an angular-momentum channel other than the one in which the resonance lies, the $3/2^+$, and also examine whether a second, background, level in the $3/2^+$ channel improves the $R$-matrix result. Fourth, we note that de Souza {\em et al.}'s additional point-to-point errors were absolute in the $S$ factor, i.e. they are given as additions (or subtractions) to $S$.
We suggest an alternative statistical model, in which the additional error is fractional in the $dt$ cross section. And we test this model, and our eventual fit to the data, by examining the residuals of that fit and seeing if they are consistent with the statistical assumptions underlying our treatment of the experimental errors.

The paper is structured as follows. In Section~\ref{sec:Rmatrix} we describe the aspects of $R$-matrix theory pertinent to this study. Section~\ref{sec:Bayes} reviews Bayes' theorem, discusses the likelihood employed to connect data to the $R$-matrix model, and lists and explains the priors on $R$-matrix parameters. It also briefly describes our MCMC sampling package and strategy. (More details on the MCMC aspects of our work are given in Appendices~\ref{sec:chain_details}, \ref{sec:sampler_diagnostics}, and \ref{sec:alt_methods}.)
Section~\ref{sec:experiments} describes the experiments and discusses the model we adopt for experimental errors, including the priors on the corresponding nuisance parameters. Section~\ref{sec:correlations} derives the correlations that arise between several $R$-matrix parameters sampled in de Souza {\em et al.}'s analysis. Understanding those correlations allows us to eliminate redundant parameters; we  display and discuss the posterior for the remaining parameters in Sec.~\ref{sec:desouza-like}. 
Section~\ref{sec:Kobzev} examines the data set of Ref.~\cite{Kobzev1966} in detail and concludes there is a systematic problem with the data that cannot be remedied without additional knowledge.
In Section~\ref{sec:final} we then present the parameter posterior for our multi-level $R$-matrix analysis, and our final evaluation of the $S$ factor. We summarize and provide an outlook in Sec.~\ref{sec:conclusion}. 
 
\section{{\em R}-Matrix Theory}
\label{sec:Rmatrix}

In the $R$-matrix approach nuclear wave functions are described inside the channel radii by many-body basis functions; outside the channel radii, they are described by a linear combination of two-body Coulomb functions. The projection of a basis function on to a particular (two-body) channel configuration at the channel radius is its reduced width amplitude. This description assumes that nuclear interactions beyond the channel radii may be neglected and that channels involving three or more nuclei can be ignored. Here, we further assume the basis functions are eigenfunctions of the Hamiltonian satisfying specified boundary conditions at the channel radii~\cite{Wig47,LaneThomas1958}. 

Observables can be calculated in the $R$-matrix approach as long as one has knowledge of the energy eigenvalues and reduced width amplitudes that summarize the dynamics inside the channel radius. In this sense one does not need the full calculation of the interior, but only those parameters.
In a phenomenological $R$-matrix analysis, the energy eigenvalues and reduced width amplitudes are treated as adjustable parameters. In such a calculation it is necessary to truncate the number of levels and channels in order to have a tractable number of unknown parameters. But it should always be borne in mind that---modulo its underlying assumptions stated above---the $R$-matrix method should converge to the full result if enough levels and channels are included in the calculation. 

The elements of the $\bm{R}$~matrix are defined by
\begin{equation} \label{eq:rmatrix}
    R_{c^\prime c} = \sum_{\lambda} \frac{\gamma_{\lambda c^\prime}\gamma_{\lambda c}}{E_\lambda-E}~,
\end{equation}
where $E_\lambda$ and $\gamma_{\lambda c}$ are the level energies and reduced-width amplitudes, respectively.
Here, the index $\lambda$ labels the levels and $c$ (or $c'$) the channels.
Channels are defined as unique combinations of particle pair type $\alpha$, orbital angular momentum $\ell$, channel spin $s$, and total angular momentum $J$.
The scattering (or collision) matrix, $\bm{\mathcal S}$, is then given by
\begin{equation}
    \bm{\mathcal S} = 2i\bm{\rho}^{1/2}\bm{O}^{-1} \left[\bm{1}-\bm{R}(\bm{L}-\bm{B})\right]^{-1} \bm{R}\bm{\rho}^{1/2}\bm{O}^{-1} + \bm{I}\bm{O}^{-1}~,
\end{equation}
with definitions of $\bm{\rho}$, $\bm{I}$, $\bm{O}$, $\bm{L}$, and $\bm{B}$ given in Ref.~\cite[Eq.~(3)]{Brune2002}. The conservation of total angular momentum and parity implies that the $\bm{R}$ and $\bm{\mathcal S}$ matrices are block diagonal with respect to $J^\pi$. The cross section for reaction channels ($\alpha\neq\alpha'$) is given by~\cite[X.3, Eq.~(3.4), p.~301]{LaneThomas1958}
\begin{equation}
\sigma_{\alpha\alpha'}=\frac{\pi}{k_\alpha^2} \sum_{J\ell\ell'ss'} \frac{2J+1}{(2j_{\alpha 1}+1)(2j_{\alpha 2}+1)} |{\mathcal S}^J_{\alpha's'\ell',\alpha s\ell}|^2 ,
\end{equation}
where $k_\alpha$ is the incoming wave number and $j_{\alpha 1}$ and $j_{\alpha 2}$ are the intrinsic spins of the nuclei making up the incoming pair $\alpha$.

Since we will be performing phenomenological analyses covering a limited range of energy, the level expansion given by Eq.~(\ref{eq:rmatrix}) must be truncated. A background level at a much higher energy will be used to represent the strength supplied by the omitted levels. The calculation of the $\bm{\mathcal S}$~matrix involves the Coulomb functions evaluated at the channel radii, the radii beyond which the nuclear interactions are assumed to vanish, and so it is not immediately obvious that observables will be independent of these channel radii. In fact, the background level plays a critical role in allowing the $\bm{\mathcal S}$~matrix to be approximately invariant as the channel radii are modified over a reasonable range~\cite{Hale:2013ama}. Since the proper value of the channel radii are somewhat ambiguous, a range of values should be investigated, in order to ensure the conclusions do not strongly depend upon the channel radius. One might think that one can modify the channel radius to values that are much larger than the distance beyond which the nuclei cease to interact, absorbing the difference in the evaluated $\bm{\mathcal S}$-matrix in the background level(s). However, very large channel radii produce artificial energy dependence in the cross section that can only be cancelled by background level(s) at lower energies, ultimately rendering an analysis with a single background level impossible and the calculation inefficient. On the other hand, nuclear interactions do not completely vanish beyond the channel radii used in phenomenological analyses. The reduced-width amplitudes should thus be interpreted as renormalized quantities, i.e., they have to absorb the effects of the nuclear interaction beyond the channel radius. If those effects become too large the calculation again becomes inefficient. Thus the best strategy is to use channel radii that cover a reasonable range at and not too far beyond the sum of the radii of the nuclei involved in the collision. Further discussion of the channel radius is available in Ref.~\cite[Sec.~IV.F]{deB17}.

In this work, we consider the $^3{\rm H}(d,n)^4{\rm He}$ reaction for center-of-mass energies below 250~keV. In this energy regime the cross section is dominated by a very strong $J^\pi=3/2^+$ resonance, which is formed with $\ell=0$ in the entrance channel. While this resonance alone may be sufficient to describe the reaction, we also consider the possibility of a non-resonant contribution (background level) with $J^\pi=1/2^+$, which can also be formed with $\ell=0$. The role a background level with $J^\pi=3/2^+$ is also investigated. The contributions of higher partial waves are smaller still: \textcite{Bem97} measured angular asymmetries of $< 1$\% in this energy domain. We therefore neglect higher partial waves in this study. 

With the limitation to $J^\pi = 1/2^+$ or~$3/2^+$ and $\ell=0$ in the entrance (${}^3{\rm H}+d$) channel, we also have $s=J$ here. In the exit (${}^4{\rm He}+n$) channel, we have $\ell=0$ ($J=1/2$) or $\ell=2$ ($J=3/2$) and $s=1/2$. For a given $J$, the channels may thus be labeled unambiguously by $d$ and $n$ for the entrance and exit channels, respectively. The particle-pair labels on the cross section, wave number, etc\ldots will be dropped when there is no ambiguity. The cross section for the $^3{\rm H}(d,n)^4{\rm He}$ reaction may now be written as
\begin{equation}
\sigma = \frac{\pi}{3k^2}\left(|{\mathcal S}_{dn}^{1/2}|^2+2|{\mathcal S}_{dn}^{3/2}|^2\right).
\end{equation}
The astrophysical $S$ factor is related to the cross section via
\begin{equation}
S = \sigma \, E \, e^{2\pi\eta}, 
\end{equation}
where $E=\frac{(\hbar k)^2}{2\mu}$ is the ${}^3{\rm H}$-$d$ center-of-mass energy, $\mu$ is the ${}^3{\rm H}$-$d$ reduced mass, and $\eta$ is the ${}^3{\rm H}$-$d$ Coulomb parameter.

If there is only a single level for a particular $J^\pi$, the square of the $d$-$n$ $\bm{\mathcal S}$-matrix element for that $J$ becomes
\begin{equation}
	\label{eq:S_matrix_one_level}
	|{\mathcal S}_{dn}^J|^2 = \frac{\hat{\Gamma}_d\hat{\Gamma}_n}{(E_0+\Delta-E)^2 +(\hat{\Gamma}/2)^2}~,
\end{equation}
with
\begin{align}
\hat{\Gamma} &=\hat{\Gamma}_d+\hat{\Gamma}_n~, \quad\quad \hat{\Gamma}_c = 2\gamma_c^2P_c~, \label{eq:Gammas} \\
\Delta &= \Delta_d + \Delta_n ~\mbox{,~and} \quad
	\Delta_c = -\gamma_c^2(S_c-B_c)~.
	\label{eq:level_shift}
\end{align}
Here $P_c$, $S_c$, and $B_c$ are the penetration
factors, shift factors, and boundary condition constants, respectively. The $\hat{\Gamma}_c$ defined by \eqref{eq:Gammas} are {\it formal} partial widths. The penetration and shift factors are given by~\cite{LaneThomas1958}:
\begin{eqnarray}
P_c(\eta_c,k_c a_c)&=& \frac{k_c a_c}{F_l^2(\eta,k_c a_c) + G_l^2(\eta,k_c a_c)} \\
S_c(\eta_c,k_c a_c)&=& P_c(\eta_c,k_c a_c) (F_l(\eta,k_c a_c) F_l'(\eta,k_c a_c) \nonumber\\
& & \qquad + G_l(\eta,k_c a_c) G_l'(\eta,k_c a_c)),
\end{eqnarray}
where the $'$ indicates differentiation with respect to $k_c a_c$, and $F_l$ and $G_l$ denote the regular and irregular Coulomb functions. We note that $\eta_n=0$, so hereafter we simply write $\eta_d \equiv \eta$.
If one further assumes that the level is a distant background level, as we do for $J^\pi=1/2^+$, i.e., $E_0 + \Delta \gg E$, the denominator of the $\bm{\mathcal S}$-matrix becomes approximately energy independent. To leading order the energy dependence of this denominator can be neglected, leading to an expression in which the $\bm{\mathcal S}$-matrix only depends on $E_0$, $\gamma_d$, and $\gamma_n$ through a single combination $A$:
\begin{equation}
    \label{eq:1hp_background}
    |{\mathcal S}^{1/2}_{dn}|^2 = \frac{4}{\pi} A_{1/2} P_{l=0}(\eta, k_da_d) P_{l=0}(0, k_na_n)~,
\end{equation}

The ${\mathcal S}$~matrix---and hence all physical observables---are independent of the choice of the boundary condition constants $B_c$, even if the number of levels is finite~\cite{Mor72,Barker72,Brune2002}. The transformation of $E_\lambda$ and $\gamma_{\lambda c}$ which ensure this as $B_c$ changes is given by \textcite{Barker72}. Since physics is independent of $B_c$ we make the convenient choice
\begin{equation}
B_c = S_c(E_\lambda)
\end{equation}
for a particular level $\lambda$. In this case, $\Delta$, as defined by Eq.~(\ref{eq:level_shift}), vanishes at $E_\lambda$ and the cross section has a maximum in the vicinity of this energy. One may then interpret this particular $E_\lambda$ as a resonance energy $E_r$.

When there are two or more levels for a given $J^\pi$,  as is the case for some of our fits to the $3/2^+$ contribution to the $^3{\rm H}(d,n)^4{\rm He}$ reaction, it is advantageous to use the alternative level energies and reduced width amplitudes defined by \textcite{Brune2002}. This approach is mathematically equivalent to the standard $R$-matrix parameterization~\cite{Wig47,LaneThomas1958}, but allows all of the parameters to be directly interpreted in terms of resonance energies and partial widths. The relationship between these parameters and the $\bm{\mathcal S}$~matrix is given in Ref.~\cite[Eqs.~(33) and~(34)]{Brune2002}. In this parameterization, there are no boundary-condition constants $B_c$.

Lastly, we note that \textcite{deSouza:2019pmr} allow for the modification of the cross section at very low energies due to the screening of the inter-nuclear Coulomb interaction by electrons in the target molecules. Following Refs.~\cite{Ass87,Eng88}, they replace:
\begin{equation}
    S(E)\rightarrow e^{\pi\eta(U_e/E)} S(E)~.
\end{equation}
This introduces another parameter into the physical description of the reaction, the electron screening potential, $U_e$. $U_e$ depends on the chemical form of the target and has been estimated to be $\approx 20-40$~eV for positive hydrogen ions incident on diatomic hydrogen gas~\cite{Lan89}. Below we refer to $U_e$ as an $R$-matrix parameter even though strictly speaking electron-screening effects are a separate issue from the treatment of the inter-nuclear Coulomb and strong forces using the $R$-matrix formalism. 

\section{Bayesian Statistics}

\label{sec:Bayes}

In a phenomenological $R$-matrix analysis the level energy and reduced width parameters must be estimated from one or more experimental data sets. Denoting the parameters collectively as $\theta$ and the data sets as $D$ our goal is to compute the posterior probability distribution $	\mathcal{P}\equiv p(\theta|D,I)$, where $I$ is other information used in the analysis, e.g., the number of channels and levels included in the analysis and the channel radii selected. Bayes' theorem relates this posterior to the likelihood $\mathcal{L} \equiv p(D|\theta,I)$ and the prior $p(\theta|I)$, according to:
\begin{equation}
	\mathcal{P} = \frac{p(D|\theta,I)~p(\theta|I)}{p(D|I)} \propto p(D|\theta,I)~p(\theta|I)~.
\end{equation}
For the purposes of this paper the factor in the denominator, $p(D|I)$ is a constant, and we can focus solely on the numerator. 

\subsection{Likelihood}

In many $R$-matrix analyses, and in ours too, the parameters $\theta$ include not just $R$-matrix parameters such as reduced width amplitudes and level energies, but also what we will term ``statistical parameters'', e.g., normalization factors that account for common-mode errors in data sets. We will also follow Ref.~\cite{deSouza:2019pmr} and allow for the possibility that point-to-point uncertainties were underestimated. We consider $N_{\rm expt}$ data sets $\{D_1, \ldots, D_j\}$ where each data set consists of $N_j$ measurements of the ${}^3{\rm H}(d,n){}^4{\rm He}$ cross section, with a corresponding error bar, $\sigma_{i,j} \pm \delta_{i,j}$ taken at a nominal energy $E_{i,j}$ (with $i=1, \ldots, N_j$ indexing the measurements in the $j$th data set). If we assign to the experiment a normalization factor $f_j$ and an additional (fractional) point-to-point uncertainty $\alpha_j$ then the likelihood takes the standard product form for independent measurements:
\begin{widetext}
\begin{equation}
    \label{eq:likelihood}
    \mathcal{L} \equiv p(D|\theta_R,{\bf f},\mathbf{\alpha},I)=\prod_{j=1}^{N_{\rm expt}} \prod_{i=1}^{N_j} \left[\frac{1}{(2 \pi (\delta_{i,j}^2 + \alpha^2_j \sigma_{i,j}^2))^{1/2}} \exp\left(-\frac{1}{2} \frac{(\sigma_{i,j} - f_j \sigma_R(E_{i,j};\theta_R))^2}{\delta_{i,j}^2 + \alpha_j^2 \sigma_{i,j}^2}\right)  \right].
\end{equation}
\end{widetext}
But, we now have additional parameters that allow for common-mode and enhanced ``statistical'' (point-to-point) errors in each data set. These parameters are listed as the vectors ${\bf f}=\{f_1, \ldots, f_{N_{\rm expt}}\}$ (for common-mode error) and $\mathbf{\alpha}=\{\alpha_1,\ldots,\alpha_{N_{\rm expt}}\}$ for point-to-point error. We will refer to these unreported point-to-point errors as ``extrinsic'' errors in keeping with the terminology of \cite{de_Souza_2020}. Meanwhile $\sigma_R(E;\theta_R)$ is the $R$-matrix result for the cross section at energy $E$, evaluated at particular values $\theta_R$ of the $R$-matrix parameters.

\subsection{Priors}

Below we present results for several $R$-matrix and statistical models. Each of the parameters in these models has a prior. The first model we work with is intentionally chosen to be very similar to that used by de Souza {\em et al.}~\cite{deSouza:2019pmr}.
The prior distributions taken for our initial model are (for more details on the physical meaning of these parameters, see Sec.~\ref{sec:Rmatrix}):
\begin{align}
    E_r & \sim U(0.020~{\rm keV}, 0.100~\rm{keV})~, \\
    \gamma_{d}^2 & \sim T(0,\infty)N(0, 3~\rm{MeV}^2)~, \\
    \gamma_n^2 & \sim T(0, \infty)N(0, 3~\rm{MeV}^2)~, \\
    U_e & \sim T(0, \infty)N(0, 1~\rm{keV}^2)~, \\
    \delta_{j,\rm{extr}} & \sim T(0, \infty)N(0, 2~\rm{b}^2)~, \\
    f_j & \sim T(0,\infty)N(1, \delta_{f,j}^2)~.
\end{align}
Here a uniform distribution in $x$ is
\begin{equation}
    U(a,b) =
    \begin{cases}
        \frac{1}{b-a} & a \leq x \leq b~, \\
        0  & x~\rm{otherwise}~,
  \end{cases}
\end{equation}
a normal distribution in $x$ is
\begin{equation}
    N(\mu,\sigma^2) = \frac{1}{\sqrt{2\pi}\sigma}e^{-\frac{(x-\mu)^2}{2\sigma^2}}~,
\end{equation}
while the truncation function, $T$, suppresses values outside of the interval defined by its arguments.

The widths of the truncated normal distributions for $f_j$ are fixed according the the systematic uncertainties reported with the data sets.
They are given in the last column of Table~\ref{tab:experiments}.
As discussed in Sec.~\ref{sec:experiments}, the Brown data set did not report an independent determination of its absolute normalization, so a large width was assigned to its prior. The other experiments all have small values of $\delta_{f,j}$, so for them the Gaussian prior on $f_j$ is effectively the same as the log-normal prior used in Ref.~\cite{deSouza:2019pmr}.

The priors on other (statistical and $R$-matrix) parameters introduced in later variants of our analysis will be stated as those parameters are introduced.

\subsection{Tools}

The addition of the statistical parameters $\mathbf{\alpha}$ and $\mathbf{f}$ means that, even in a relatively simple reaction like ${}^3{\rm H}(d,n){}^4{\rm He}$, we need to determine the posterior 	$p(\theta|D,I)$ in a space of dimensionality as high as 17. The only efficient way to sample the posterior is via MCMC sampling.

In this work we employ the package \texttt{emcee} \cite{Foreman_Mackey_2013} which implements an affine-invariant ensemble sampler. The sampler deploys many simultaneous, inter-dependent walkers to explore the provided posterior and allows the practitioner full control over their physical, statistical, and data models.

We now discuss some details of our MCMC sampling, in 
the interest of future reproducibility. 
While the standard parameters of \texttt{emcee} generically work well, we found that in many cases, tuning some of the sampler parameters can significantly improve the efficiency of the sampler. 
Specifically, in the language of \texttt{emcee}, proposals are generated by ``moves''.
By default, these moves are Goodman \& Weare \cite{goodman2010} ``stretch'' moves. Empirically, we found a mixture of differential evolution (20\%) and snooker proposals (80\%) reduces autocorrelation times appreciably.

If we treat the channel radii $a_d$ and $a_n$ as part of the parameter set $\theta_R$ then, 
as we will show in Section~\ref{sec:correlations}, the posterior is extended and involves non-linear correlations. 
Several attempts were made to overcome the long autocorrelation times induced by these correlations.
Sampling techniques and reparameterizations used in those attempts are discussed briefly in Appendix~\ref{sec:alt_methods}.

\section{Experimental Data}
\label{sec:experiments}

\begin{table}
    \centering
    \begin{tabular}{|c|c|c|c|c|c|}
        $j$ & Reference & Year & Energy Range & $N_j$ & $\delta_{f,\,j}$ \\
        &&& (keV) && \\
        \hline
        5 & \textcite{Con52} & 1952 &  $12\le E\le 214$ & 43 & 0.030 \\
        4 & \textcite{Arn53} & 1953 & $9\le E\le 70$ & 53 & 0.020 \\
        3 & \textcite{Kobzev1966} & 1966 & $46\le E\le 264$ & 45 & 0.025 \\
        1 & \textcite{Jar84} & 1984 & $5\le E\le 47$ & 17 & 0.0126 \\
        2 & \textcite{Bro87} & 1987 & $48\le E\le 70$ & 8 & 0.5 
    \end{tabular}
    \caption{Summary of the ${}^3{\rm H}(d,n){}^4{\rm He}$ cross section data sets analyzed. The columns provide the data set number $j$, reference, year, energy range, number of points included $N_j$, and relative systematic uncertainties $\delta_{f,\,j}$. The data set numbers chosen in \cite{deSouza:2019pmr} are replicated in this work.}
    \label{tab:experiments}
\end{table}

The data $D$ used to obtain the posterior includes ${}^3{\rm H}(d,n){}^4{\rm He}$ cross sections from Refs.~\cite{Con52,Arn53,Kobzev1966,Jar84,Bro87}, the same five sources used by \textcite{deSouza:2019pmr}. We agree with their data selection criteria, as other sources have much larger uncertainties and/or are higher in energy. We use the same data points from each reference as were employed in \textcite{deSouza:2019pmr}, although some details in the treatment of data differ, as discussed below. A summary of the data sets is given in Table~\ref{tab:experiments}.

\subsection{\texorpdfstring{\textcite{Con52}}{Conner {\em et al.}} (1952)}

We take 43 of their reported $90^\circ$ differential cross section measurements with $21\le E_d\le 357$~keV, where $E_d$ is the deuteron laboratory energy, corresponding to $12\le E\le 214$~keV. Following \textcite{deSouza:2019pmr}, we do not include the two lowest energy points which are subject to larger uncertainties. The differential cross sections are converted to total cross sections by converting to the c.m. system and multiplying by $4\pi$. The correction from converting to the c.m. system is small, but not negligible, amounting to a 1.6\% increase in the cross section for the highest energy we consider. We note that \textcite{deSouza:2019pmr} did not convert to the c.m. system. Following \textcite{deSouza:2019pmr}, we assume a 1\% point-to-point uncertainty for this data set. The original paper is not clear on this point, as on page 471 it states ``The statistical probable error from each target was about 1 percent'' but on page 472 it states ``The probable error in the number of counts is less than $\frac{1}{2}$ percent.'' \textcite{Con52} do not provide a detailed discussion of systematic errors in their cross sections, but they do estimate the probable error to be ``about 2\%.'' Converting the probable error to a standard deviation assuming a Gaussian distribution, we adopt 3.0\% for the normalization uncertainty. This value is significantly larger that the 1.8\% assumed by \textcite{deSouza:2019pmr}. \textcite{Con52} do not make a clear statement about their energy uncertainties.

\subsection{\texorpdfstring{\textcite{Arn53}}{Arnold {\em et al.}} (1953)}

We utilize 53 of their reported total cross sections with $15\le E_d\le 117$~keV, corresponding to $9\le E\le 70$~keV. Following \textcite{deSouza:2019pmr}, we do not include points with $7\le E_d\le 11$~keV, due to larger uncertainties, and four other points with factor of 10 clerical errors are not included. The available description of the experiment is very detailed and systematic errors are carefully considered and well controlled. The measurement was conducted by measuring $\alpha$ particles at $90^\circ$ in the laboratory; the correction for c.m. motion in the determination of the total cross section was taken into account. We utilize the same point-to-point uncertainties as \textcite{deSouza:2019pmr}, which are based on Table~VIII of the original paper. The absolute systematic normalization uncertainty of 2.0\% adopted by \textcite{deSouza:2019pmr} appears reasonable and is also adopted here. The uncertainty in the c.m. energy is given in Table~VIII of the original paper as 0.18~keV at $E=15$~keV, 0.17~keV at $E=30$~keV, and 0.077~keV at $E=60$~keV.

\subsection{\texorpdfstring{\textcite{Kobzev1966}}{Kobzev {\em et al.}} (1966)}

We employ 45 of their reported $90^\circ$ differential cross section measurements with $115\le E_t\le 660$~keV, where $E_t$ is the triton laboratory energy, corresponding to $46\le E\le 264$~keV. The differential cross sections are converted to total cross sections by converting to the c.m. system and multiplying by $4\pi$. The correction from converting to the c.m. system is significant, amounting to a 4.7\% increase in the cross section for the highest energy we consider. We again note that---as with the \textcite{Con52} data set---\textcite{deSouza:2019pmr} did not convert to the c.m. system. The publication~\cite{Kobzev1966} supplies little information regarding experimental details or systematic errors. We follow \textcite{deSouza:2019pmr} and assume a point-to-point uncertainty varying with energy between 2.0\% and 2.5\%, and a 2.5\% systematic uncertainty in the absolute cross section. These experimental data are also subject to rather large uncertainties in the energy, with the uncertainty quoted to be 2.5\% for measurements with $E\le 60$ and 2.0\% for $60\le E\le 480$~keV. 
The energy uncertainty is most probably dominated by the calibration of the magnetic analyzer and the energy loss corrections, which would make them highly correlated in energy. Evidence for an energy-dependent systematic error in this data set is presented below in Sec.~\ref{sec:Kobzev}.

\subsection{\texorpdfstring{\textcite{Jar84}}{\em Jarmie {\em et al.}} (1984)}

We consider their 17 reported total cross sections covering $12.5\le E_t\le 117$~keV, corresponding to $5\le E\le 47$~keV. This experiment took extensive steps to minimize systematic uncertainties and the publication~\cite{Jar84} provides considerable documentation of those uncertainties and of the experiment in general. The differential cross section for $\alpha$ particles was measured at six laboratory angles, which were converted the c.m. system. The differential data were consistent with c.m. isotropy and converted to a total cross section by averaging and multiplying by $4\pi$. We and \textcite{deSouza:2019pmr} adopt the quoted point-to-point uncertainties and absolute normalization uncertainty of 1.26\%. The systematic error in energy varies from 0.048\% at the lowest energy to 0.014\% at the highest energy, while the random uncertainty in the energy calibration varies from 0.008\% at the lowest energy to 0.004\% at the highest energy.

\subsection{\texorpdfstring{\textcite{Bro87}}{Brown {\em et al.}} (1987)}

We use their 8 reported total cross sections covering $80\le E_d\le 116$~keV, corresponding to $48\le E\le 70$~keV. These data were taken using the same equipment as the experiment reported in Ref.~\cite{Jar84}, but with the role of beam and target interchanged. In this case, the absolute target density was not determined, leaving the absolute cross section scale undetermined. The data are therefore treated as relative measurements. Otherwise, the methods of data reduction are as in Ref.~\cite{Jar84}. We and \textcite{deSouza:2019pmr} utilize the quoted point-to-point uncetainties. The uncertainty in the c.m. energy is 9~eV.



    

\section{Understanding the correlations between {\em R}-matrix parameters that emerge from sampling}

\label{sec:correlations}

 \textcite{deSouza:2019pmr} combines a sophisticated Bayesian model with a single-level, two-channel $R$-matrix parameterization. In addition to the parameters identified above---$E_0$, $\gamma_d^2$, and $\gamma_n^2$, $U_e$---de Souza {\em et al.}  choose to also indirectly sample the boundary-condition parameter $B$. Although, instead of sampling and reporting $B$ they parameterize their model in terms of the energy, $E_B$, at which the level shift $\Delta_c$ is zero, i.e., $E_B$ and $B$ are related by $B=S_c(E_B)$.
As already mentioned, de Souza {\em et al.} also consider the channel radii, $a_d$ and $a_n$, as parameters to be sampled. 
The de Souza {\em et al.} $R$-matrix parameter set, $\theta_{R, {\rm de S}}$, is thus
\begin{equation} 
    \label{eq:theta_ds}
	\theta_{R,{\rm deS}} \equiv \{E_0, E_B, \gamma_d^2, \gamma_n^2,a_d,a_n,U_e\}~.
\end{equation}
In this section we identify one redundancy and three correlations inherent to this version of the $R$-matrix parameterization. We  eliminate the redundancy, thereby reducing the dimensionality of the $\theta_R$ space we are sampling. We also discuss how we deal with complications introduced by the correlation of channel radii with the reduced channel widths. 

We observe significant $E_0$--$E_B$,  $\gamma_d^2$--$a_d$,  $\gamma_n^2$--$a_n$, and  $\gamma_d^2$-$\gamma_n^2$ correlations when we sample the posterior $p(\theta_{R, \rm{deS}},{\bf f},\mathbf{\alpha}|D,I)$, see Figs.~\ref{fig:e0_eb_correlation}, \ref{fig:gc2_ac_correlation}, and \ref{fig:ds_rpars}.
These correlations---especially the last two---lead to untenably large autocorrelation times: as much as an order of magnitude larger than those reported in \cite{deSouza:2019pmr}. 

\subsection{\texorpdfstring{$E_0$}{E0}--\texorpdfstring{$E_B$}{EB} correlation}

\begin{figure}
    \centering
    \includegraphics{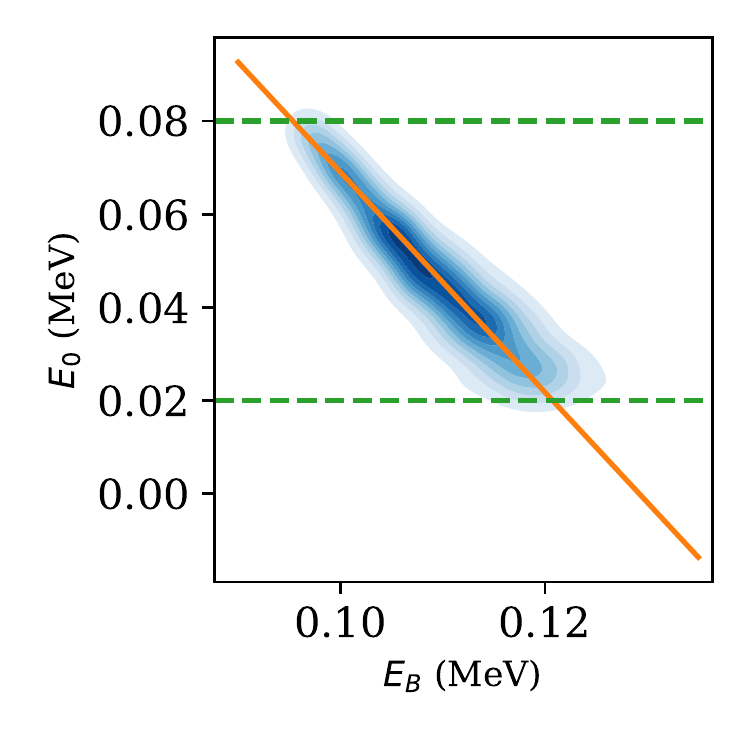}
    \caption{Correlation between $E_0$ and $E_B$. Blue, shaded regions are the samples from the MCMC analysis. The solid, orange line is the predicted correlation according to Eq.~\eqref{eq:E0EB}. The green, dashed lines are the boundaries of the uniform prior applied to $E_0$ in Ref.~\cite{deSouza:2019pmr}.}
    \label{fig:e0_eb_correlation}
\end{figure}

It has been shown by Barker \cite{Barker72} that level energies and reduced widths can be ``renormalized'' as the boundary condition is changed:
there exists a relationship between $E_B$ and $E_0$ that leaves physical observables invariant~\cite{Barker72}. 
This relationship presents itself as a correlation in the multi-dimensional posterior for $\theta_{R,{\rm deS}}$, see
Fig.~\ref{fig:e0_eb_correlation}.

To derive an analytic description of the relationship we start from  \eqref{eq:S_matrix_one_level}.
The first term in the denominator of Eq.~\eqref{eq:S_matrix_one_level} is
\begin{equation}
    \label{eq:e0_eb_term}
    E_0 + \Delta_d + \Delta_n - E~,
\end{equation}
where the $E_B$ prescription means that $\Delta_c = -\gamma_c^2\left(S_c - S_c(E_B)\right)$.
For a single-level parametrization, the combination $E_0 + \Delta_d + \Delta_n$ determines the location (in $E$) of the peak of the resonance.
For this case, we know from \textcite{Barker72} that different boundary condition parameters, $B_c=S_c(E_B)$, will reproduce the same cross section provided that $E_0$ is suitably adjusted.
In light of that, we take the derivative of (\ref{eq:e0_eb_term}) with respect to $E_B$ and set it to zero.
Solving for $\frac{dE_0}{dE_B}$ gives
\begin{equation}
    \frac{dE_0}{dE_B} = -\gamma_d^2\left.\frac{dS_d}{dE}\right|_{E=E_B} - \gamma_n^2 \left. \frac{dS_n}{dE}\right|_{E=E_B}~.
    \label{eq:E0EB}
\end{equation} 
Assuming that the derivatives on the right-hand side do not vary significantly with $E_B$ yields the result that the $E_0$--$E_B$ correlation will be a straight line whose slope is given by the right-hand side of \eqref{eq:E0EB}.
The comparison with a set of MCMC samples is shown in \ref{fig:e0_eb_correlation}.
The agreement between the predicted slope and the samples is excellent.
The $E_0$ prior taken in Ref.~\cite{deSouza:2019pmr} was a uniform distribution from 20 keV to 80 keV.
Those bounds are shown in the Fig.~\ref{fig:e0_eb_correlation}. Eq.~\eqref{eq:E0EB} maps this information on $E_0$ into what is apparently a tight constraint on $E_B$, cf. Ref.~\cite[Table I]{deSouza:2019pmr}. In fact, the correlation between $E_0$ and $E_B$ extends much further than this in both directions.

\subsection{The correlation between \texorpdfstring{$a_c$}{ac} and \texorpdfstring{$\gamma_c^2$}{gammac2}}

The single-level $R$-matrix formulation for the cross section may be mapped into the traditional Breit-Wigner formula if one chooses $B_c=S_c(E_r)$.
If one further performs a first-order Taylor expansion of $S_c$ around the resonance energy, $E_r$, the resulting partial widths are
\begin{equation}
    \Gamma_c = \frac{2\gamma_c^2P_c}{1+\sum_{c^\prime}\gamma_{c^\prime}^2\frac{dS_{c^\prime}}{dE}}~,
    \label{eq:bw_partial_widths}
\end{equation}
where $P_c$ and the energy derivative of the shift functions are evaluated at $E_r$. If the channel radii are varied and the $R$-matrix parameters $E_r$ and $\gamma_c^2$ re-optimized at new values of $a_c$, it is reasonable to expect the $\Gamma_c$ will remain constant.
This condition should govern how $\gamma_c^2$ depends on $a_c$. 

We note that 
these ``Breit-Wigner'' or ``observed'' partial widths differ from the formal partial widths of $R$-matrix theory\cite{LaneThomas1958} by the volume renormalization factor $1+\sum_{c^\prime}\gamma_{c^\prime}^2\frac{dS_{c^\prime}}{dE}$. Because of this term
$\Gamma_c$ depends on all of the $\gamma_c^2$'s.
Typically, the volume renormalization factor is close to unity.
However, for the low-energy $3/2^{+}$ resonance in the $^3$H$(d,n)^4$He reaction, this is not the case: for the $R$-matrix parameters found by Barker~\cite{Barker97} it is approximately 6, with the dominant contribution coming from the deuteron channel.
Nevertheless, as shown in \cite{Azuma2010}, Eq.~\eqref{eq:bw_partial_widths} may be inverted to yield
\begin{equation}
    \gamma_c^2 = \frac{\Gamma_c}{P_c} \left[2 - \sum_{c^\prime}\frac{\Gamma_{c^\prime}}{P_{c^\prime}}\frac{dS_{c^\prime}}{dE}\right]^{-1}~.
    \label{eq:inv_bw_partial_widths}
\end{equation}

Imposing invariance of $\Gamma_c$ as $a_c$ is changed leads to a relationship between $\gamma_c^2$ and $a_c$ that can be parameterized as a power law
\begin{equation}
    \label{eq:ag2_corr}
    a_c^{\alpha_c}\gamma_c^2 = \beta_c~,
\end{equation}
where $\alpha_c$ and $\beta_c$ are extracted from solutions of Eq.~\eqref{eq:inv_bw_partial_widths}. These solutions suggest $\alpha_n=1.25$ and $\beta_n=0.352$; $\alpha_d=4.28$ and $\beta_d=6970$. 

\begin{figure}
    \centering
    \includegraphics{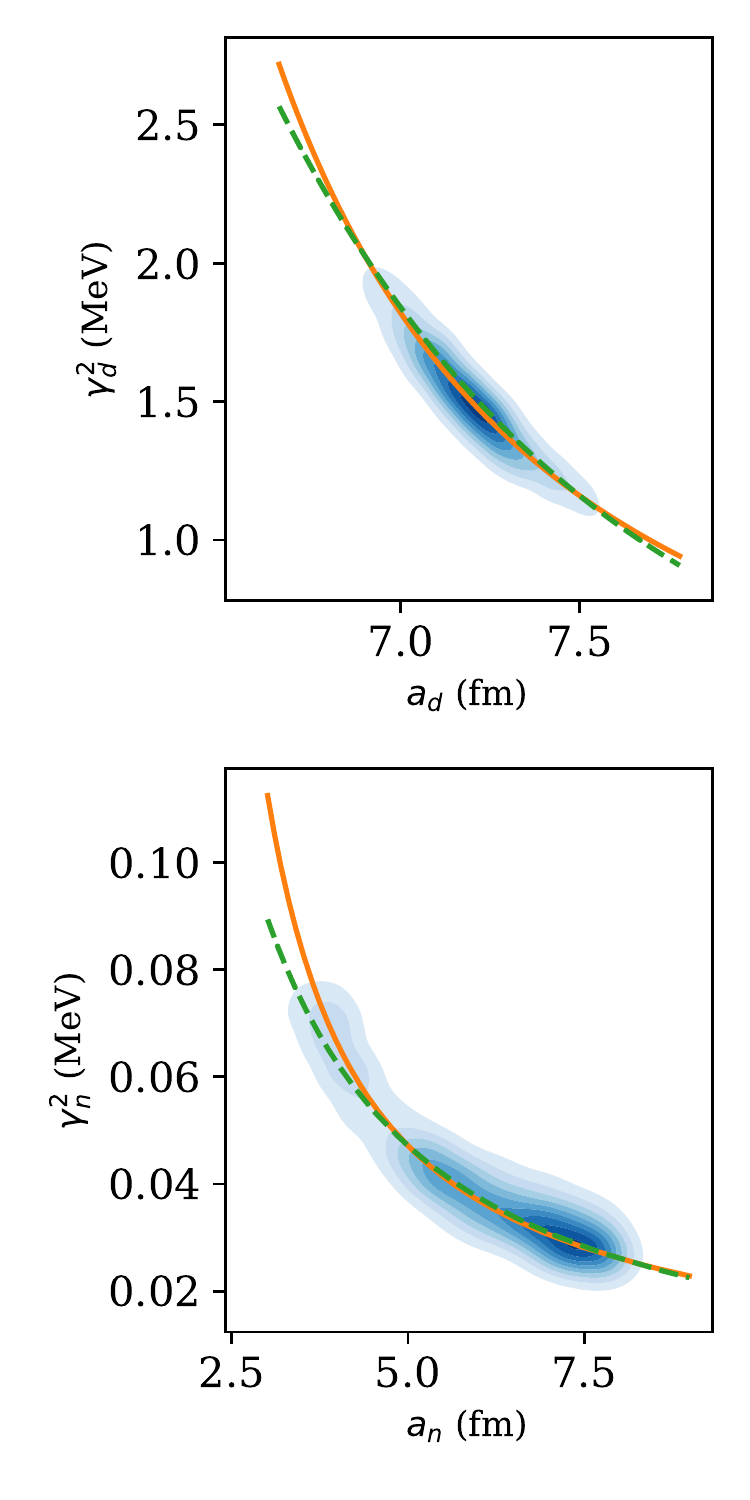}
    \caption{Correlations between $\gamma_d^2$ and $a_d$ (top) and $\gamma_n^2$ and $a_n$ (bottom). Samples from the MCMC analysis are shown as blue circles. Solid, orange lines represent the correlation imposed by Eq.~\eqref{eq:inv_bw_partial_widths}. Dashed, green lines represent the fitted results of Eq.~\eqref{eq:ag2_corr}.}
    \label{fig:gc2_ac_correlation}
\end{figure}

We compare these semi-analytical predictions to the sampling results
for both deuteron and neutron channels in Fig.~\ref{fig:gc2_ac_correlation}. The agreement between the samples and the two predicted results is excellent.

The nonlinearity and extended nature of these $a_c$-$\gamma_c^2$ correlations is largely responsible for the extremely long autocorrelation times we saw when sampling with the parameter set which includes the radii that was defined in \textcite{deSouza:2019pmr}. For more discussion of possible ways to overcome those correlations see Appendix~\ref{sec:alt_methods}.

\subsection{\texorpdfstring{$\gamma_d^2$}{gammad2}--\texorpdfstring{$\gamma_n^2$}{gamman2} correlation}
\label{sec:unitarity}

\begin{figure}
    \centering
    \includegraphics[width=\linewidth]{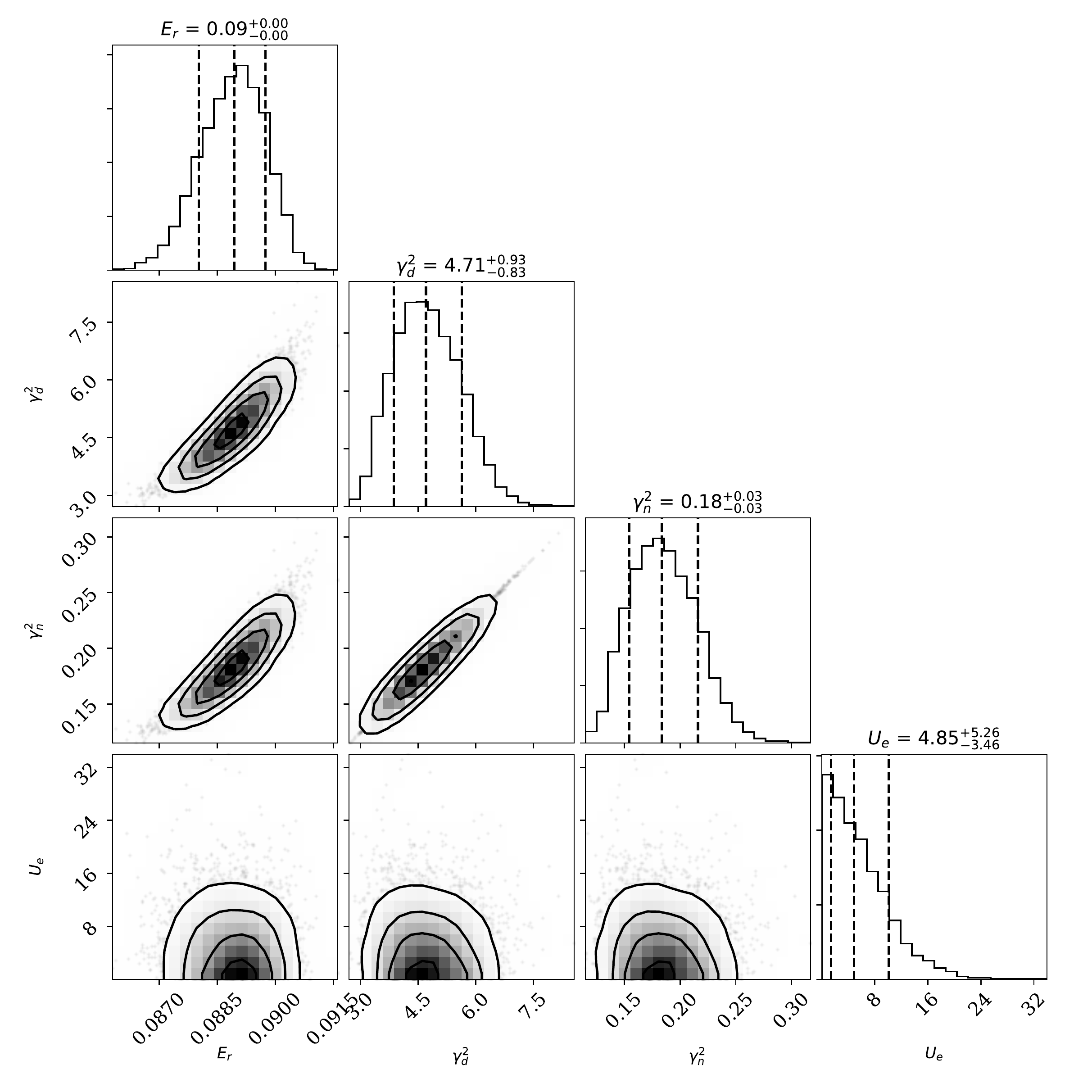}
    \caption{$R$-matrix-parameter posteriors for a single-level $R$-matrix analysis at $a_d=5.56$ fm and $a_n=3.633$ fm combined with statistical model $\mathcal{U}_{af}$.}
    \label{fig:ds_rpars}
\end{figure}

Finally, we discuss a correlation that is very clear in the $R$-matrix parameter posterior Fig.~\ref{fig:ds_rpars}, but---unlike the correlations discussed so far in this section---represents physics and not parameter redundancy. The correlation between $\gamma_d^2$ and $\gamma_n^2$ is a consequence of the fact that the cross section for $dt$ fusion at low energies almost saturates the maximum value allowed by quantum-mechanical unitarity, when $|{\mathcal S}_{dn}^J|^2=1$~\cite{Barker97}.

In the single-level formulation, Eq.~(\ref{eq:S_matrix_one_level}) shows that this limit is achieved at the resonance energy if $\Gamma_d=\Gamma_n$. 
The height of a single resonance peak is always proportional to $\Gamma_d \Gamma_n$ and so this product is well constrained by data. The sum $\Gamma=\Gamma_d+\Gamma_n$ is also well determined, since it is the total width of the resonance in the Breit-Wigner formula. This makes it seem as if $\Gamma_d$ and $\Gamma_n$ can be independently determined, but in fact if $\Gamma_d$ and $\Gamma_n$ are approximately equal the predicted cross section is insensitive to the individual values of $\Gamma_d$ or $\Gamma_n$. This can be seen by fixing $\Gamma$, in which case the product $\Gamma_d\Gamma_n=\Gamma_d(\Gamma-\Gamma_d)$. The derivative of this product with respect to $\Gamma_d$ then vanishes when $\Gamma_d=\Gamma/2$. 
It follows that if we are close to the unitarity limit (where $\Gamma_d=\Gamma_n$) changes to the individual widths do not affect the cross section to first order, as long as $\Gamma_d + \Gamma_n$ stays fixed. Even though the unitarity limit is not fully realized here, the final values of $\Gamma_d$ and $\Gamma_n$ put us close enough to it that the parameters $\gamma_d^2$ and $\gamma_n^2$ end up significantly correlated.

\subsection{Implications for Bayesian {\em R}-matrix practice}

In light of the correlations and redundancies induced when $E_B$, $a_d$, and $a_n$ are included as parameters to be estimated, we make two decisions that reduce the dimensionality of the space in which we are sampling. 
First, we characterize the level energy and boundary condition simultaneously by requiring that $E_0 = E_B\equiv E_r$. Since the $E_B$ parameter is redundant this has no impact on our analysis. In subsequent analyses the authors of Ref.~\cite{deSouza:2019pmr} also eliminated the $E_B$ parameter and did not sample it, see, e.g., Ref.~\cite{de_Souza_2020}.

Second, we fix the channel radii to obtain the results that follow. This eliminates the non-linear correlations  seen in Figs.~\ref{fig:gc2_ac_correlation}.  We conduct our analysis at several different ($a_d$,$a_n$) pairs to ensure that physical observables predicted by our analysis are indeed insensitive to the channel radius. 


\section{Results for a one-level {\em R}-matrix model}
\label{sec:desouza-like}

Given the correlations discussed in the previous section we will formulate our comparison with \textcite{deSouza:2019pmr} with a reduced $R$-matrix parameter set:
\begin{equation}
    \theta_R \equiv \{E_r, \gamma_d^2, \gamma_n^2, U_e\}~,
\end{equation}
where, as in \cite{deSouza:2019pmr}, we include an electron-screening potential parameter, $U_e$, and use the Lane and Thomas \cite{LaneThomas1958} parameterization. The resulting $R$-matrix model is physically equivalent to the model of de Souza {\em et al.}, permitting an accurate comparison between the results shown in this section and the results reported in \cite{deSouza:2019pmr}.
But our parameterization is simpler, which allows us to readily explore different statistical models.
Perhaps the most significant difference between the approaches is that we do not sample the channel radii, but instead consider a grid of at least nine points in channel radius space: $a_d$ is fixed at 4.25, 5.56, and 7.25 fm and $a_n$ is fixed at 3.633, 5.5, and 7.5 fm.

\subsection{The statistical model of de Souza et al.}
\label{sec:ds_stat_model}

In this section we consider two different statistical models for the point-to-point errors.
The simplest model takes the errors ``as they come'' from the original publications.
The more sophisticated model, as implemented by \cite{deSouza:2019pmr}, adds an overall additional extrinsic error to each data set. 
To be precise, we will refer to the statistical model defined by
\begin{equation}
    \label{eq:Uf}
	S_{{\rm exp},i} \sim N(f_S S(E_{{\rm exp},i})~, \delta_{S,i,stat}^2)
\end{equation}
as statistical model $\mathcal{U}_f$, 
\begin{equation}
    \label{eq:Uaf}
    S_{\rm{exp}} \sim N(f_S S(E_{{\rm exp}, i}), \delta_{S,\rm{extr}}^2+\delta_{S,i, stat}^2)
\end{equation}
as statistical model $\mathcal{U}_{af}$, and
\begin{equation}
    \label{eq:Ua}
    S_{\rm{exp}} \sim N( S(E_{{\rm exp}, i}), \delta_{S,\rm{extr}}^2+\delta_{S,i, stat}^2)
\end{equation}
as statistical model $\mathcal{U}_a$.
Both models include multiplicative normalization factors, $f_S$, applied to the theory prediction, $S(E_i)$. (Note that in this subsection we treat the $S$ factor as the dependent variable, so as to perform as close a comparison as possible to \cite{deSouza:2019pmr}, 
In the sections that follow we treat the cross section as the dependent variable, as indicated in Eq.~\eqref{eq:likelihood}. This is an important change because the cross section---which is, after all, what is actually measured---has strong energy dependence at low energies.)

\begin{figure}
    \centering
    \includegraphics{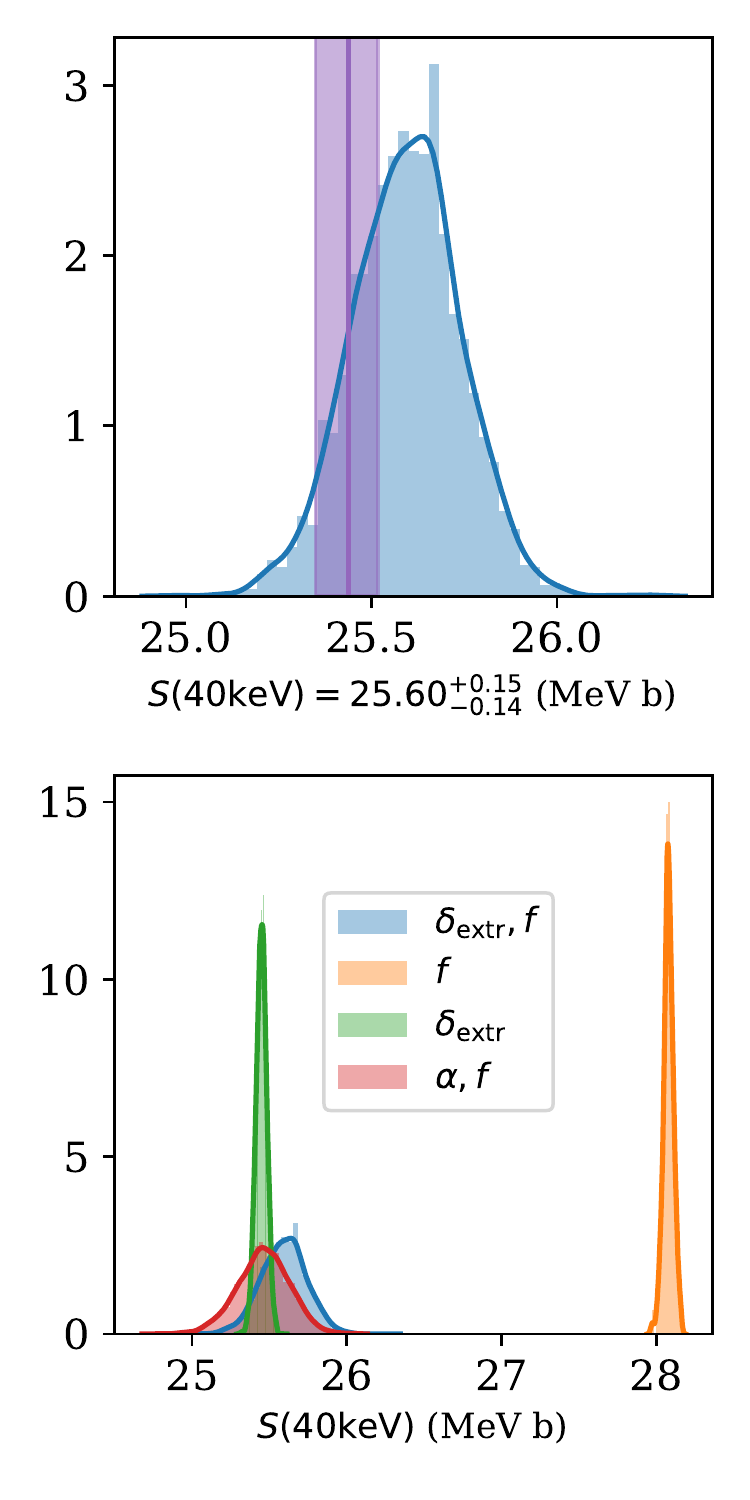}
    \caption{$S(40~\rm{keV})$ posteriors for different statistical models. The $S(40~\rm{keV})$ posteriors were generated from an analysis with a single $R$-matrix level at fixed channel radii $a_d=5.56$ fm and $a_n=3.633$ fm. $\mathcal{U}_{af}$ (blue) is shown in the top panel in comparison to a summary of the result of \cite{deSouza:2019pmr} (purple). The posteriors for four different statistical models, $\mathcal{U}_{af}$ (blue), $\mathcal{U}_f$ (orange), $\mathcal{U}_{a}$ (green), and $\mathcal{U}_{rf}$ (red), are presented in the bottom panel for comparison.}
    \label{fig:ds_s40}
\end{figure}

To demonstrate the importance of sampling extrinsic uncertainties and normalization factors, four posteriors are shown in bottom plot of Fig.~\ref{fig:ds_s40}.
Results with $\mathcal{U}_{af}$ are shown in blue and those with $\mathcal{U}_{f}$ are shown in orange.
The drastic difference in width and central value highlights the importance of sampling extrinsic uncertainties.
Without them, the reported $S$ factor is significantly larger and narrower.
The $\ln\mathcal{L}$ values for $\mathcal{U}_{af}$ are three orders of magnitude larger than those obtained using $\mathcal{U}_f$.
The improvement in likelihood is expected, as inflating the error bars necessarily improves the quality of the fit. But the dramatic improvement seen here suggests that the errors quoted in some of the original papers drastically underestimate the point-to-point uncertainty. 
Normalization factors alone are not enough to overcome the discrepancies between the data sets. And it is not surprising that the point-to-point uncertainties in the early experiments need to be inflated, as older studies tend not to include careful quantification of this type of uncertainty. For example, the measurement of \textcite{Arn53}, which is the best documented of the early experiments, only gives information about the point-to-point uncertainty from counting statistics. This uncertainty is very small (0.2\%-0.3\%) and other random or pseudo-random effects likely make significant contributions.

Included in Fig.~\ref{fig:ds_s40} is the $S(40~\rm{keV})$ posterior obtained when absolute extrinsic uncertainties are sampled without normalization factors, model $\mathcal{U}_{a}$.
The $\ln\mathcal{L}$ values obtained with $\mathcal{U}_a$ are three orders of magnitude larger than those obtained using $\mathcal{U}_f$.
This makes the point that extrinsic uncertainties carry the weight in this analysis when resolving discrepant data. 
But the bottom panel of Fig.~\ref{fig:ds_s40} also shows that if normalization uncertainties are neglected the final posterior for evaluated quantities is overly narrow.


The  $S(40~{\rm keV})$ posterior obtained with a single-level $R$-matrix and statistical model $\mathcal{U}_{af}$ at fixed channel radii $a_d=5.56$ fm and $a_n=3.633$ fm is also shown in the top panel of Figure \ref{fig:ds_s40}, where it is compared to the posterior of Ref.~\cite{deSouza:2019pmr}.
We obtain $S(40~\rm{keV}) = {25.60}^{+0.15}_{-0.14} ~\rm{MeV~b}$ in this analysis, where they reported ${25.438}^{+0.080}_{-0.089} ~\rm{MeV~b}$.
While the posteriors overlap at one standard deviation, we do not reproduce the result of \textcite{deSouza:2019pmr}, despite having used their median channel radii.
The most significant difference is the widths of the distributions.
A detailed comparison follows, but the most striking difference between the two analyses is that the posteriors  for the $R$-matrix parameters presented in Ref.~\cite{deSouza:2019pmr} do not exhibit the extended correlations described and derived in the previous section.
Such correlations should be present; we found them using a variety of sampling strategies.

\begin{figure}
    \centering
    \includegraphics[width=\linewidth]{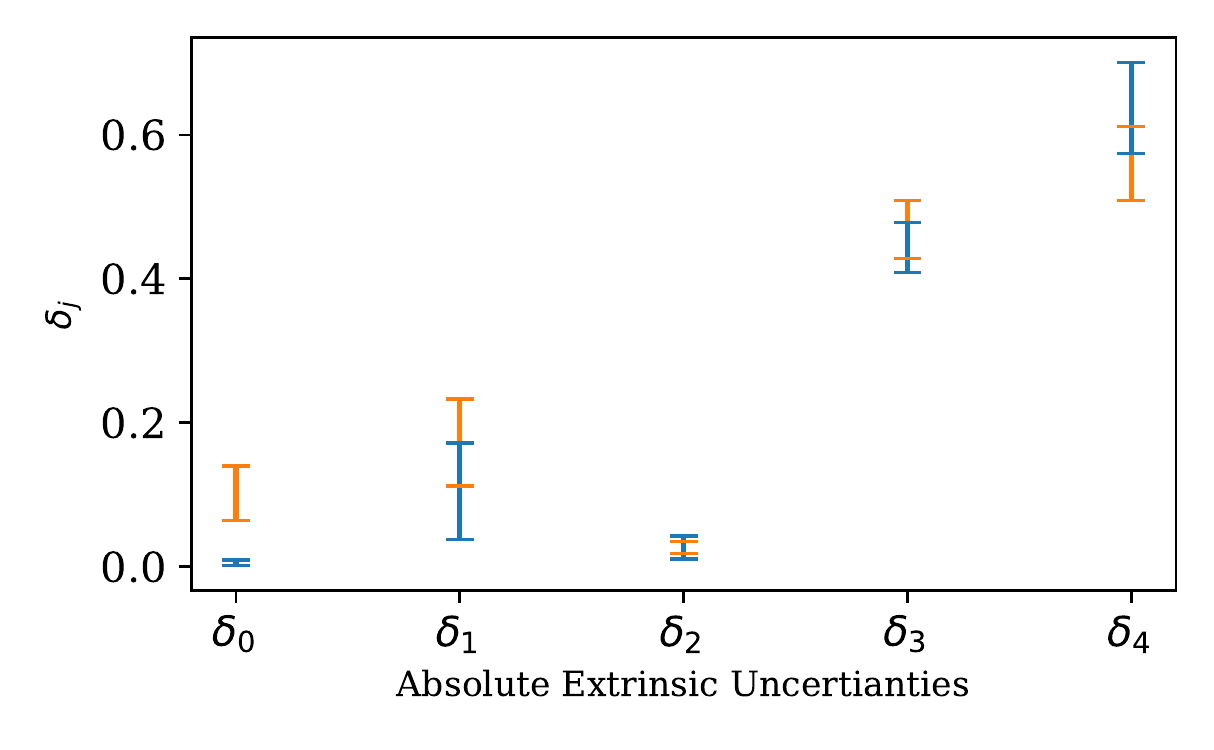}
    \caption{Summary of the absolute extrinsic uncertainty posteriors from an analysis using a single $R$-matrix level and channel radii of $a_d=5.56$ and $a_n=3.633$ fm. The lower and upper limits of the blue error bars correspond to 16\% and 84\% quantiles, respectively. For comparison, the same quantiles reported in \cite{deSouza:2019pmr} are shown in orange.}
    \label{fig:ds_delta_compare}
\end{figure}

A summary of the posterior pdf for the absolute extrinsic uncertainties $\mathbf{\delta}_{j,{\rm extr}}$ is shown in Fig.~\ref{fig:ds_delta_compare}.
While our posterior for the $R$-matrix parameters has notable differences from that of Ref.~\cite{deSouza:2019pmr},
we find absolute extrinsic uncertainties that are mostly consistent with the medians given in \cite{deSouza:2019pmr}---although
our result for the extrinsic uncertainty associated with the \textcite{Jar84} data, $\delta_{1, \rm{extr}}$, is much smaller.

\begin{figure}
    \centering
    \includegraphics[width=\linewidth]{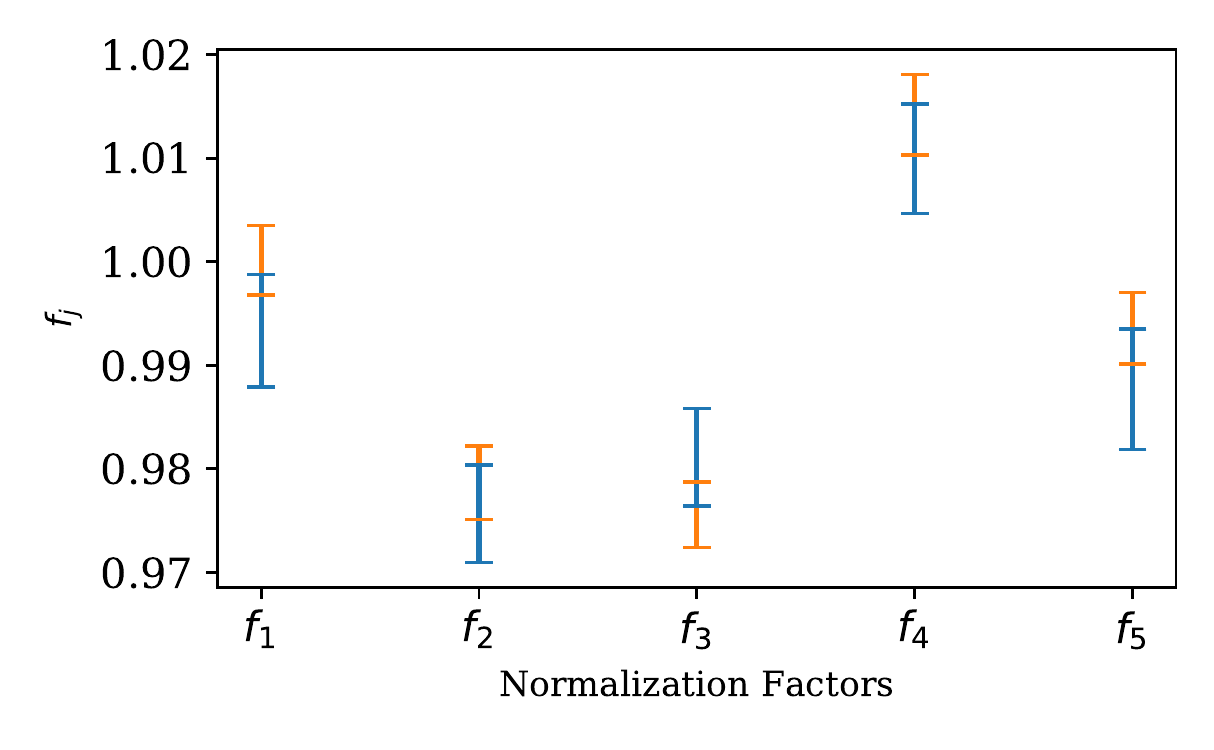}
    \caption{Comparison of the normalization factors obtained with a single-level $R$-matrix approximation and absolute extrinsic uncertainties. Our results are shown in blue. The results reported in \cite{deSouza:2019pmr} are shown in orange. The error bars indicate 16\% and 84\% quantiles.}
    \label{fig:ds_f_compare}
\end{figure}

Importantly, all of our results for normalization factors, $f_j$, have error bars that are nearly a factor of two larger than those of \textcite{deSouza:2019pmr}, as shown in Figure \ref{fig:ds_f_compare}.
This feeds directly into the comparison of $S(40~{\rm keV})$, as uncertainty in the normalization factors directly impacts the width of $S(40~\rm{keV})$ as seen in Fig.~\ref{fig:ds_s40}.
Our 16\% and 84\% quantiles both fall 0.14 MeV b from the median value of 25.60 MeV b.
Our $S(40~{\rm keV})$ posterior is nearly twice as wide as that of \textcite{deSouza:2019pmr} and our median lies just above the 95\% bound of their distribution.
The different normalization uncertainties we obtain compared to \textcite{deSouza:2019pmr} drive this different evaluation of $S(40~{\rm keV})$. 

\subsection{A different statistical model}

\label{sec:diff_stat_model}

Absolute extrinsic uncertainties can add unnecessarily large error bars at both high energies---where the cross section, and the corresponding $S$ factor, decreases significantly---and low energies, where Coulomb suppression renders the cross section exponentially small. Considering that backgrounds in the measurements are very small, it also seems more likely that the extrinsic uncertainty should be fractional rather than absolute. One situation where a fractional uncertainty is appropriate would be if there were pseudo-random variations in the detection efficiency due to changes in the beam-target intersection. Therefore, rather than sampling an extrinsic uncertainty, $\delta_{j, \rm{extr}}$, that has a fixed size for each experiment, we now construct a new statistical model that is more appropriate to this particular reaction in which the extrinsic uncertainties for a particular experiment are a certain fraction of the observable, i.e., 
\begin{equation}
    \sigma_{\rm{exp}} \sim N\left(f_{j}~\sigma(\theta; E_{\rm{exp}}), (\alpha_j \sigma_{\rm{exp}})^2 + \delta_{j,\rm{stat}}^2\right)~,
    \label{eq:Urf}
\end{equation}
where $\alpha_j$ are relative extrinsic uncertainties.
We will refer to this model as $\mathcal{U}_{rf}$.
It is defined in terms of the cross section because from now on  the cross section is sampled rather than the $S$ factor.
In cases where we apply $\mathcal{U}_{af}$ to cross-section data, the model is defined by taking $S\rightarrow\sigma$ in Eq.~\eqref{eq:Uaf}.
Similarly, where we apply $\mathcal{U}_{rf}$ to $S$-factor data, we take $\sigma\rightarrow S$ in Eq.~\eqref{eq:Urf}. We would like to emphasize that the choice of model for the extrinsic uncertainties needs to be considered on a case-by-case basis, taking into consideration the experimental methods employed, importance and nature of backgrounds, etc...

The prior adopted for all $\alpha_j$ parameters is
\begin{equation}
    \label{eq:alpha_prior}
    \alpha_j \sim T(0, \infty)N(0, 2^2)~.
\end{equation}
This distribution is extremely wide considering the extent to which an $\alpha_j$ of even 1 would inflate the error bars. 
In practice, the posteriors for $\alpha_j$ indicate small extrinsic errors; these priors have no influence on the final values of the $\alpha_j$'s and are much wider than is necessary. However, we did not need to reduce their width in order for sampling to converge. 

We now compare the statistical models $\mathcal{U}_{rf}$ and $\mathcal{U}_{af}$---with 
both relative and absolute extrinsic uncertainties applied to the cross section. The residuals are defined by
\begin{equation}
    \label{eq:residuals}
    \mathcal{R}_{i,j} \equiv \frac{\sigma_{i,j} - f_{j*}\sigma_{R}(E_{i,j};\theta_{R*})}{\sqrt{\delta_{i,j}^2 + \left(\alpha_{j*}\sigma_{R}(E_{i,j};\theta_{R*})\right)^2}},
\end{equation}
where $\theta_{*}\equiv \{\theta_{R*},\mathbf{\alpha}_{j*}, f_{j*} \}$ are the $R$-matrix and statistical-model parameters that yield either the maximum posterior probability ($\theta_{*}^{(\mathcal{P})}$; ``MAP values'') or maximum likelihood ($\theta_{*}^{(\mathcal{L})}$). 
A comparison of the residuals of the two models at the $\max{\ln{\mathcal{L}}}$ of each reveals no statistical preference for one model over the other.
However, the models' values of $\max{\ln{\mathcal{L}}}$ differ by more than 2 with relative extrinsic uncertainties producing the higher likelihood.
This difference is accounted for by the normalization factors of the normal distributions associated with each data point (first factor in Eq.~(\ref{eq:likelihood})). Absolute extrinsic uncertainties lead to huge relative cross-section errors at low energies where the cross section gets small. The likelihood is then suppressed by the normalization factors. 

\section{What energy sampling reveals about the Kobzev data set}
\label{sec:Kobzev}

If observables calculated with a given $R$-matrix approximation are channel-radius dependent, it is an indication that the analysis is suffering from at least one of two possible defects: (1) a lack of levels or (2) a misreporting of data and/or its associated error. 
The Kobzev \cite{Kobzev1966} data set was found to induce channel radius dependence in both one- and two-level $R$-matrix approximations. We now describe how we traced this back to what appears to be an energy-dependent systematic error in the data set.

\begin{figure}
    \centering
    \includegraphics{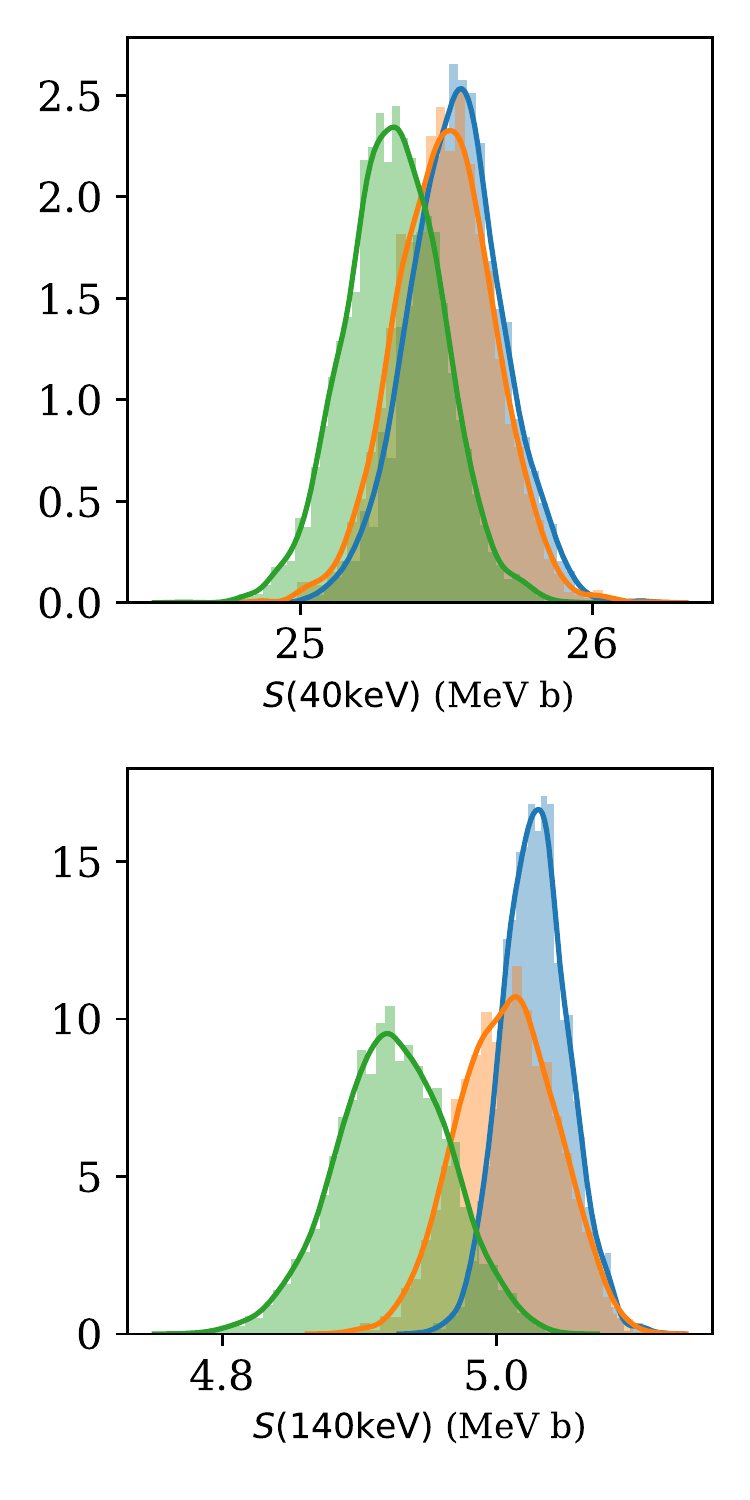}
    \caption{$S(40~\rm{keV})$ (top) and $S(140~\rm{keV})$ (bottom) posteriors using two $3/2^+$ $R$-matrix levels and statistical model $\mathcal{U}_{rf}$ for the analysis of all data sets, including that of Kobzev {\it et al.}. Posteriors are shown at $a_d=$ 4.25 (blue), 5.56 (orange), and 7.25 (green) fm. $a_n$ is fixed at 3.633 fm.}
    \label{fig:kobzev_s40_s140}
\end{figure}

We analyzed all five data sets with both our $R$-matrix ``Model A'' and ``Model B''. These models include one and two $3/2^+$ ${}^5$He levels respectively. Both include one $1/2^+$ (background) level. (For full details see Sec.~\ref{sec:final}.)
For both models we performed an analysis on a grid of 9 points in channel-radius space. 
With the Kobzev data in the likelihood the $S(40~{\rm keV})$ results obtained at different channel radii are consistent with one other within one standard deviation, see Fig.~\ref{fig:kobzev_s40_s140}. But there is a definite trend with $a_d$, as well as marked changes in the relative extrinsic uncertainty, $\alpha_3$, added in quadrature to the statistical uncertainties reported in the Kobzev set \cite{Kobzev1966}. Notably, there is a dramatic $a_d$ dependence in $\ln{\mathcal{L}}^{(K)}_A$ as well---an undesirable feature of an $R$-matrix analysis. This dependence can be seen in Table~\ref{tab:modelAB_maxlnL}.
\begin{table}
    \centering
    \begin{tabular}{|c|c|c|c|c|}
        $a_d$ (fm) & $\max\ln{\mathcal{L}}^{(K)}_A$ & $\max\ln{\mathcal{L}}^{(K)}_B$ & $\max\ln{\mathcal{L}}_A$ & $\max\ln{\mathcal{L}}_B$ \\
        \hline
        4.25 & 343.2 & 341.9 & 285.4 & 285.9 \\
        5.56 & 346.3 & 341.9 & 281.6 & 286.1 \\
        7.25 & 342.9 & 346.9 & 269.1 & 285.9
    \end{tabular}
    \caption{$\max\ln\mathcal{L}$ values models with one (A) and two (B) $3/2^+$ levels. Analyses of all five data sets including that of \textcite{Kobzev1966} are distinguished by a superscript $K$.}
    \label{tab:modelAB_maxlnL}
\end{table}

These details are manifested in the posteriors of the $S$ factor shown in Fig.~\ref{fig:kobzev_s40_s140} at 40 and 140 keV. The physical model should deliver consistent results across the entire energy range being analyzed and it does not for $S(140~{\rm keV})$.

In order to investigate these issues further we implemented energy sampling (sampling in the independent variable) for this data set. Following Ref.~\cite{deSouza:2019pmr} we constructed a likelihood for the energies of the form
\begin{equation}
    \label{eq:energy_sampling_model}
    E_{{\rm exp},i} \sim N(E_i + f_E, \delta_{{\rm stat},i}^2 + \delta_{E, {\rm extr}}^2)~,
\end{equation}
where $E_{{\rm exp},i}$ represents the $i$th reported experimental energy, $E_i$ is the energy at which the $R$-matrix cross section is to be evaluated, $f_E$ allows for a systematic shift to be sampled, $\delta_{{\rm stat},i}$ is the statistical energy uncertainty reported with the $i$th data point, and $\delta_{E,{\rm extr}}$ allows for an additional, point-to-point uncertainty to be sampled as well.

\begin{figure}
    \centering
    \includegraphics[width=\linewidth]{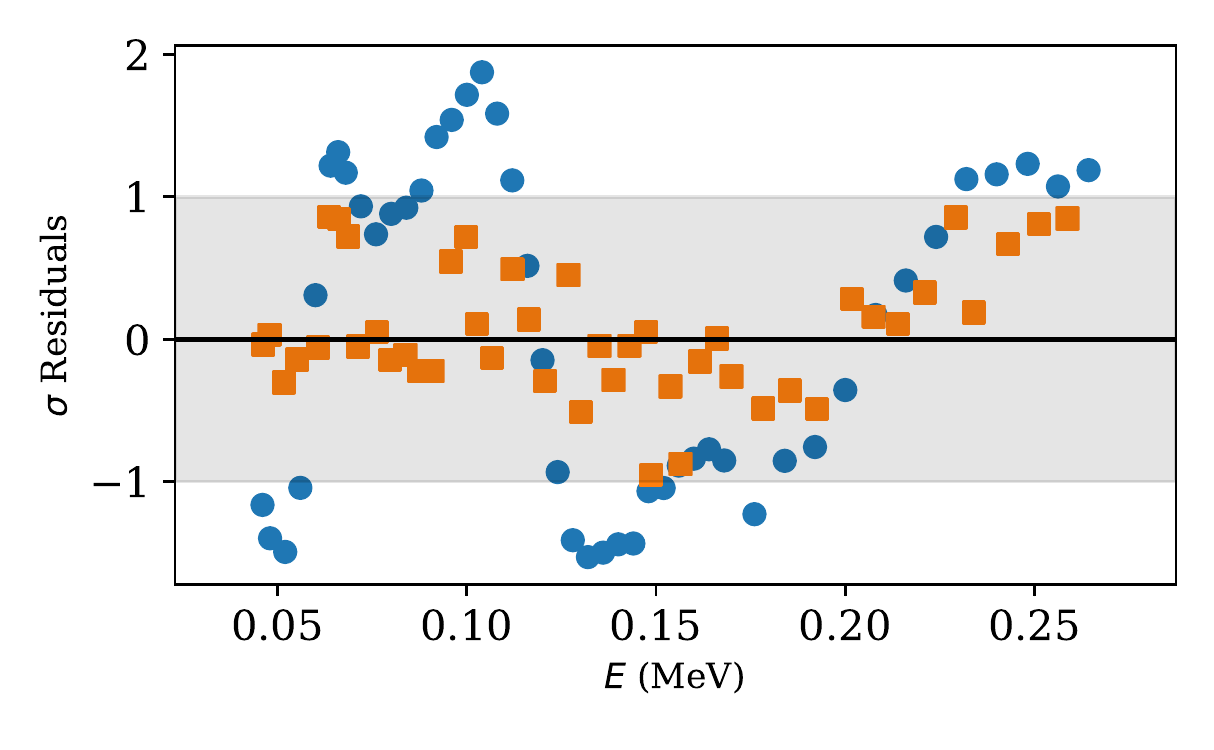}
    \caption{Cross section residual comparison of the Kobzev data when energies are sampled (orange squares) versus when they are not (blue circles).}
    \label{fig:residuals_kobzev_sampled_energies}
\end{figure}

In Figure \ref{fig:residuals_kobzev_sampled_energies}, cross section residuals are shown for two cases. First, the blue circles represent the residuals at $\theta_{*}^{(\mathcal{L})}$ without this additional energy sampling.
The residuals when the energies are sampled according to the likelihood derived from Eq.~(\ref{eq:energy_sampling_model}) are shown in orange.

The cross section residuals using the sampled energies have much better statistical properties---i.e., they are much more consistent with the hypothesis of point-to-point noise---than are the ones without. And the need for an extrinsic statistical error on the cross section, i.e., an inflation of the point-to-point errors, also is greatly reduced once energy sampling is introduced, leading to a reduction in the median value of the cross section extrinsic error by a factor of 5.
This energy-sampling analysis was only of the Kobzev {\it et al.} data set, which may account for part of this decrease, but the marked drop in $\alpha_j$ is broadly consistent with  Fig.~\ref{fig:residuals_kobzev_sampled_energies}. Energy sampling improves the internal statistical consistency of the Kobzev {\it et al.} data.
The resulting posteriors of the statistical parameters, $f_E$ and $\delta_{E,{\rm extr}}^2$, indicate only a small overall energy shift ($\approx 1.5$ keV) and no need for an inflation of the (already sizable) energy uncertainty: the additional contribution to the uncertainty of individual energies is less than 0.43 keV (84\% credibility interval). 

\begin{figure}
    \centering
    \includegraphics[width=\linewidth]{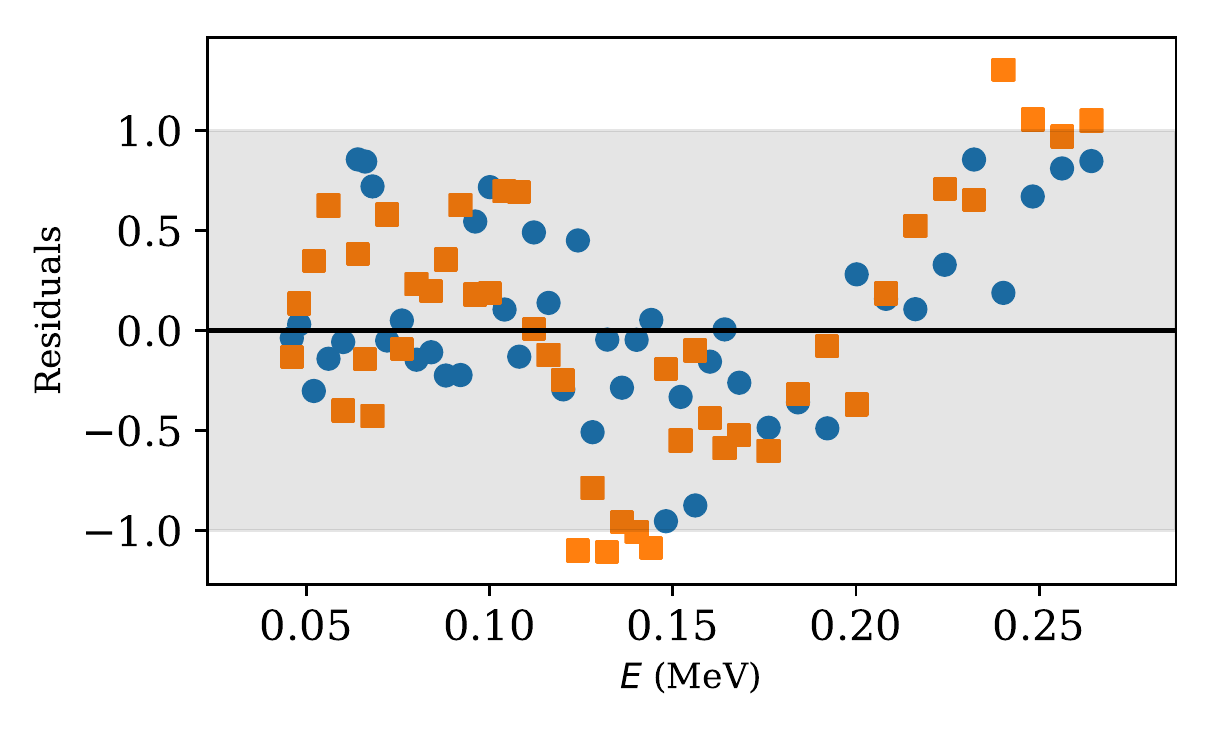}
    \caption{Residual comparison of the Kobzev data. The cross section residuals are shown as blue circles, and the energy residuals are shown as orange squares.}
    \label{fig:kobzev_residuals}
\end{figure}

But, while these parameters don't indicate large effects, sampling
the energies does reveal something peculiar in the energy range from 150--250 keV. 
Figure \ref{fig:kobzev_residuals} shows the sampled energy residuals in orange alongside the cross section residuals in blue.
The decidedly non-random behavior above $\approx 150$ keV indicates that there is a systematic misreporting of the energies in the Kobzev set that is not captured by a random error in the observed cross section or $S$ factor. We emphasize that when the energy sampling is implemented each energy $E_i$ can move independently, so it is significant that they ``choose'' to arrange themselves in this fashion between 150--250 keV. 
The cross section residuals---which should also be random---show a systematic trend there too.

\begin{figure}
    \centering
    \includegraphics[width=\linewidth]{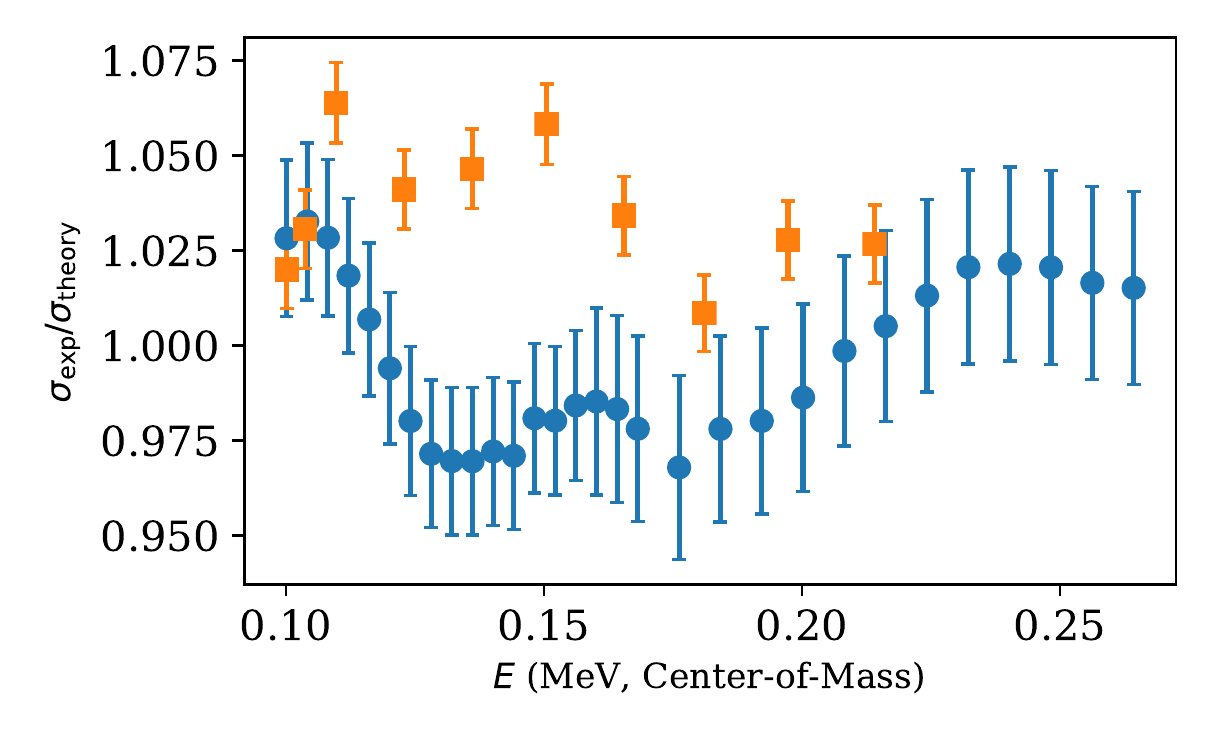}
    \caption{Experimental cross sections above 100 keV (COM) relative to the theory prediction at $\max\ln\mathcal{L}$ for the Kobzev (blue circles) and Conner (orange squares) data sets.}
    \label{fig:KobzevConnercomparison}
\end{figure}

Furthermore, above the resonance the Kobzev data set disagrees with the only other $dt$ data set that extends above 70 keV, that from Conner {\em et al.}---see Fig.~\ref{fig:KobzevConnercomparison}. Indeed, the $a_d$-dependence in the $S(140~{\rm keV})$ posterior (see bottom panel of Fig.~\ref{fig:kobzev_s40_s140}) can be traced to the presence of two solutions for the $R$-matrix model, one that agrees with the Kobzev data and one that agrees with the Conner data. Allowing the beam energies reported in the Kobzev publication to float within their reported error reveals that an energy-dependent systematic uncertainty affects the Kobzev data in the energy region 150--250 keV.  While energy sampling can correct for a drift in beam energy, doing so assumes uncorrelated energy errors, which is clearly not consistent with the pattern of energy residuals in Fig.~\ref{fig:kobzev_residuals}.
Without knowing further details of the experiment it is difficult to determine the correct statistical model for this systematic uncertainty. It is also unfortunately the case that there are no other data sets in this energy range that could shed additional light on this issue. Because of these issues we do not use the Kobzev data set in the analysis that follows. 

\section{A more sophisticated \texorpdfstring{$R$}{R}-matrix model}

\label{sec:final}

We constructed two other models that go beyond the model we used when attempting to reproduce the results of \textcite{deSouza:2019pmr} (see Sec.~\ref{sec:desouza-like}).
The first, Model A, consists of a single $3/2^+$ level and a single $1/2^+$ background level.
Model B is a further extension that also adds a $3/2^+$ background level fixed at 10 MeV.
Note that both models include the $1/2^+$ background level according to \eqref{eq:1hp_background} in an incoherent sum with the effects of the $3/2^+$ channel; 
$A_{1/2}$ is the dimensionless parameter that characterizes the contribution of this $1/2^+$ level. 
Model A and Model B are both parameterized in terms of the ``Breit-Wigner'' partial widths defined in Eq.~\eqref{eq:bw_partial_widths} and are formulated using the Brune parameterization. The $R$-matrix parameter sets for each are thus:
\begin{equation}
    \label{eq:modelA_theta}
    \theta_{R} \equiv \{E_r, \Gamma_{1d}, \Gamma_{1n}, U_e, A_{1/2} \}~,
\end{equation}
for Model A and
\begin{equation}
    \label{eq:modelB_theta}
    \theta_{R} \equiv \{E_r, \Gamma_{1d}, \Gamma_{1n}, \Gamma_{2d}, \Gamma_{2n}, U_e, A_{1/2} \}~,
\end{equation}
for Model B.
The corresponding prior distributions are
\begin{align}
    \label{eq:final_priors}
    E_r & \sim U(0.020, 0.100~\rm{MeV}) \\
    \Gamma_{ic} & \sim 
    \begin{cases}
        U(0, \Gamma_{ic,\rm{WL}}),& \Gamma_{ic} <= \Gamma_{ic,\rm{WL}}\\
        \Delta_{ic}N(\Gamma_{ic,\rm{WL}}, \Gamma_{ic,\rm{WL}}^2) & \Gamma_{ic} > \Gamma_{ic,\rm{WL}}
    \end{cases} \\
    U_e & \sim T(0,\infty)~N(0, 0.001^2~\rm{MeV}^2) \\
    A_{1/2} & \sim T(0,\infty)~N(0,1^2) \\
    \alpha_j & \sim T(0,\infty)~N(0,2^2) \\
    f_j & \sim T(0,\infty)~N(0, \delta_{j, \rm{syst}}^2)~,
\end{align}
where $i$ denotes the $R$-matrix levels 1 and 2, $j$ denotes the data sets 1 (Jarmie), 2 (Brown), 4 (Arnold), and 5 (Conner), and $c$ denotes the channels $d$ (deuteron) and $n$ (neutron).
$\Delta_{ic}$ is a factor applied to the normal distribution such that the probability density function is continuous at the boundary, $\Gamma_{ic,\rm{WL}}$.
This boundary is the approximate Wigner limit for the partial width in level $i$ and channel $c$ given by
\begin{equation}
    \label{eq:gamma_WL}
    \Gamma_{ic} = \frac{2P_{ic}\gamma_{ic}^2}{1 + \gamma_{ic}^2 \frac{dS_{c}}{dE}(E_{i})}~,
\end{equation}
where the reduced width amplitude in channel $c'\neq c$ is taken to be zero (see \eqref{eq:bw_partial_widths}).
$E_i$ is 100 keV for $i=1$ and 10 MeV for $i=2$.

While Model A’s implementation is independent of the choice of standard or Brune parameterization~\cite{Brune2002}, our implementation of Model B does have some subtleties that, in practice, make a dramatic difference in how quickly the MCMC sampling converges. 
The parameterization in terms of partial widths and the use  of the Brune parameterization both lead to significant improvements in sampling efficiency for some channel radii pairs.

\begin{figure}
    \centering
    \includegraphics[width=\linewidth]{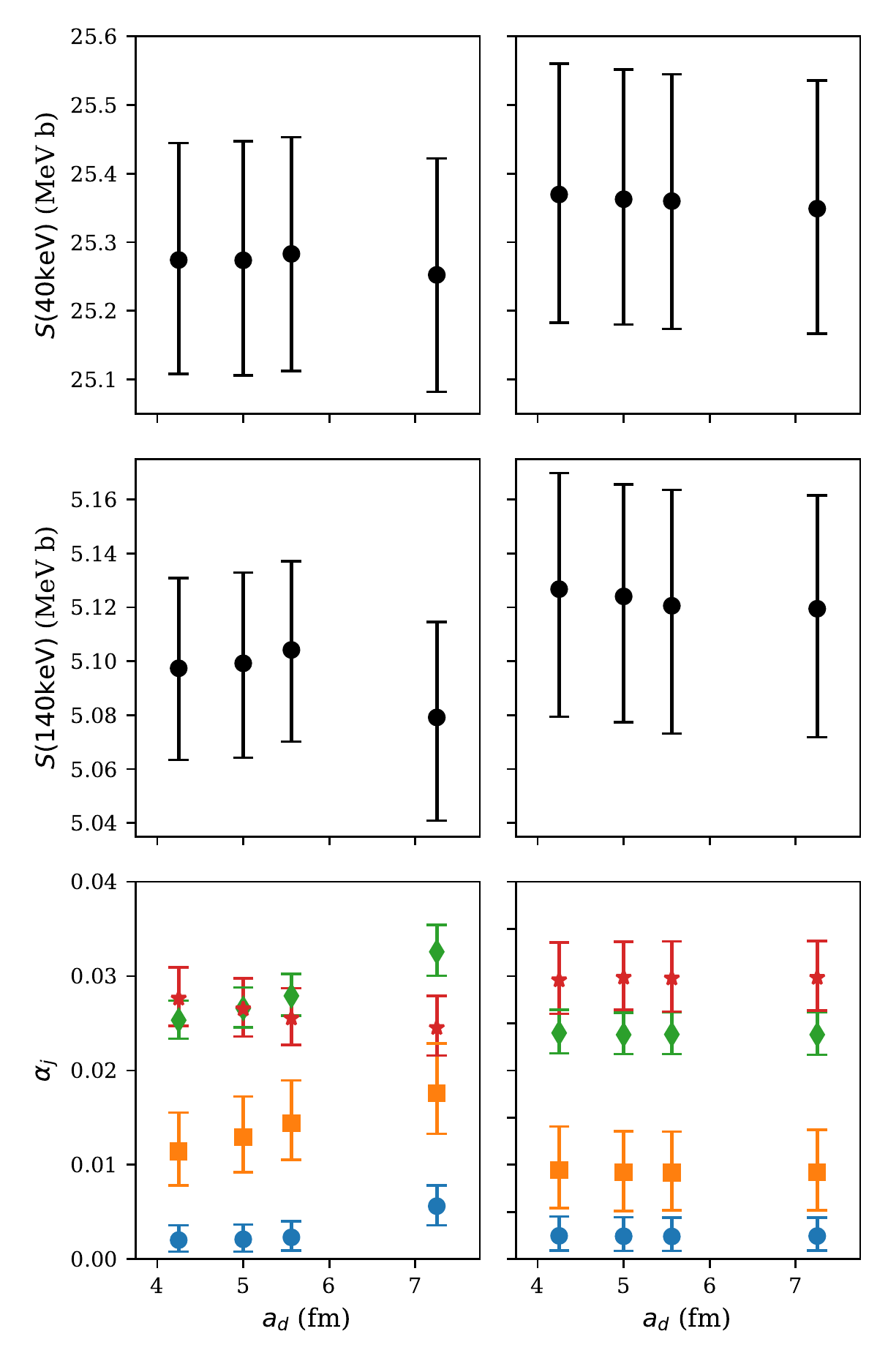}
    \caption{$S(40\rm{keV})$, $S(140\rm{keV})$, and $\alpha_j$ dependence on $a_d$. Model A results are shown in the left column. Model B results are shown in the right column. Blue circles correspond to results obtained for the \textcite{Jar84} (blue circles), \textcite{Bro87} (orange squares), \textcite{Arn53} (green diamonds), and \textcite{Con52} (red stars) are shown individually in the bottom row. Error bars reflect the 16\% and 84\% quantiles.}
    \label{fig:modelAB_ad_dependence}
\end{figure}

Turning to the results, the left column of Fig.~\ref{fig:modelAB_ad_dependence} shows that, once the Kobzev data set is excluded from the analysis, the $S$-factor and extrinsic uncertainty results in Model A are remarkably stable with respect to $a_d$. 
Here, in addition to the values of $a_d$ evaluated in the previous sections, we also include $a_d = 5.00$ fm, which was used by \textcite{Bro87}.
A significant decrease in the $\max\ln{\mathcal{L}}_A$ is observed as $a_d$ increases, as seen in Table \ref{tab:modelAB_maxlnL}.
Although this difference is not physically observable, the strong preference for smaller $a_d$ values is a difficult feature to tolerate in an $R$-matrix analysis. 
As the extrinsic uncertainties for the four remaining data sets were stable with respect to channel radii, we see no indication of hidden systematic uncertainties in the data and instead look to the addition of a $3/2^+$ background level to remove this channel-radius dependence.

The $a_d$ dependence of the Model B results is summarized in the right column of Figure \ref{fig:modelAB_ad_dependence}.
They are consistent with Model A and stable, but Model B performs quantitatively better by two measures.
First, the $\max\ln{\mathcal{L}}_B$ values are nearly $a_d$-independent (see Table \ref{tab:modelAB_maxlnL}); the variation in $\ln{\mathcal{L}}_B$ is more than an order of magnitude smaller than with Model A.
Second, all $\ln{\mathcal{L}}_B$ values are higher than the highest $\ln{\mathcal{L}}_A$ values.
Clearly, the additional 3/2$^+$ level at 10 MeV dramatically suppresses the preference for smaller deuteron channel radius that exists in Model A.

The full posterior of the $R$-matrix parameters in this model can be found in Appendix~\ref{sec:chain_details}, see Fig.~\ref{fig:blip_model_rpar_corner}. Here we elucidate two key features. 

\begin{figure}
    \centering
    \includegraphics[width=\linewidth]{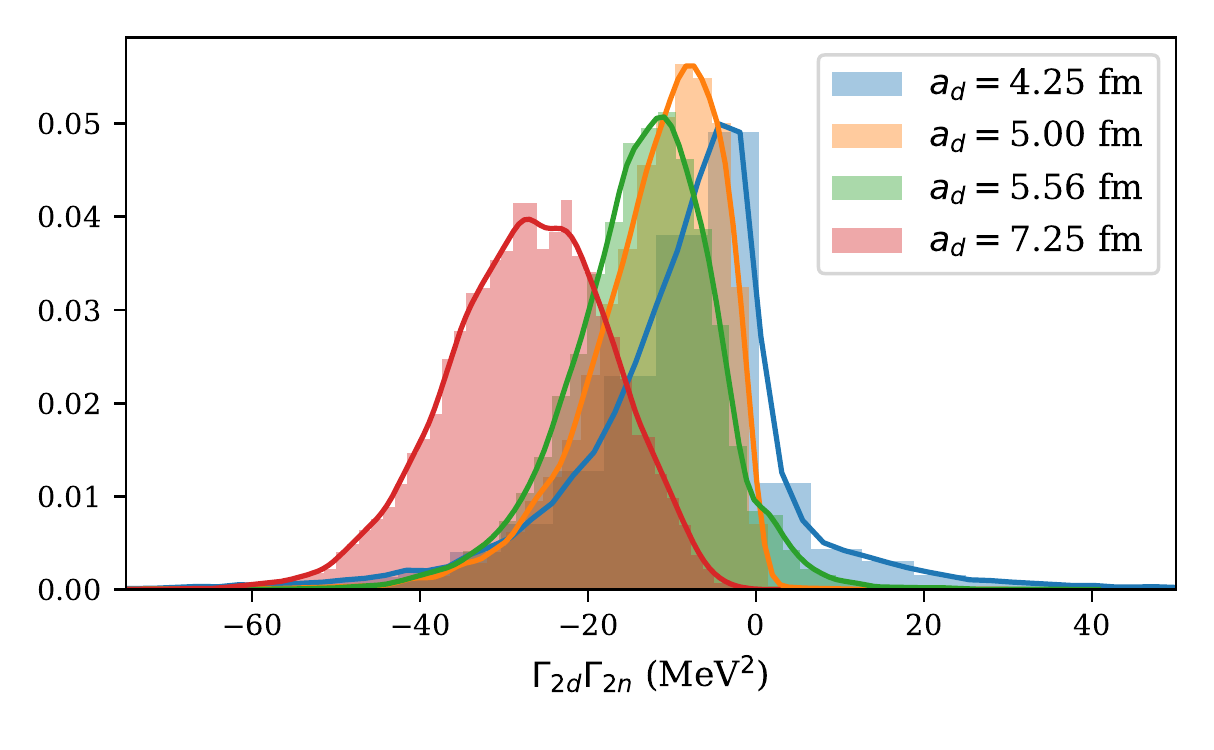}
    \caption{The evolution of the product $\Gamma_{2d}\Gamma_{2n}$ is given for all $a_d$ values under consideration. Results were obtained with $R$-matrix Model B and statistical model $\mathcal{U}_{rf}$. $a_n$ was fixed at 3.633 fm.}
    \label{fig:modelB_g2d_ad_dependence}
\end{figure}

First,
the contribution of the 3/2$^+$ background level can be assessed from the magnitude of $\Gamma_{2d}\Gamma_{2n}$.
Figure \ref{fig:modelB_g2d_ad_dependence} shows the evolution of $\Gamma_{2d}\Gamma_{2n}$ as $a_d$ increases.
There are two important trends here.
First, as $a_d$ increases, the magnitude of the peak value of $\Gamma_{2d}\Gamma_{2n}$ increases.
Second, the probability density at $\Gamma_{2d}\Gamma_{2n}=0$ \textit{tends} to decrease with increasing $a_d$ indicating that the contribution of the background level is more prominent at larger $a_d$.
This explains Model A's preference for smaller $a_d$.
As $a_d$ increases, more input is required from the background level, and Model A is simply not equipped to provide it.

Second, we note that the electron screening potential, $U_e$, and the $1/2^+$ background level both have very little impact on the final results. $U_e$ is less than $18.6$ eV (84\% credibility).
In comparison, \textcite{deSouza:2019pmr} report $U_e\le 14.7$ eV at the 97.5\% credibility level.
The $1/2^+$ fractional contribution to the $S$-factor, $S^{(1/2^+)}(E)/S(E)$ is always $< 1.3$\%, attaining its maximum value at the lowest energy considered, approximately 5 keV. 

\begin{figure}
    \centering
    \includegraphics[width=\linewidth]{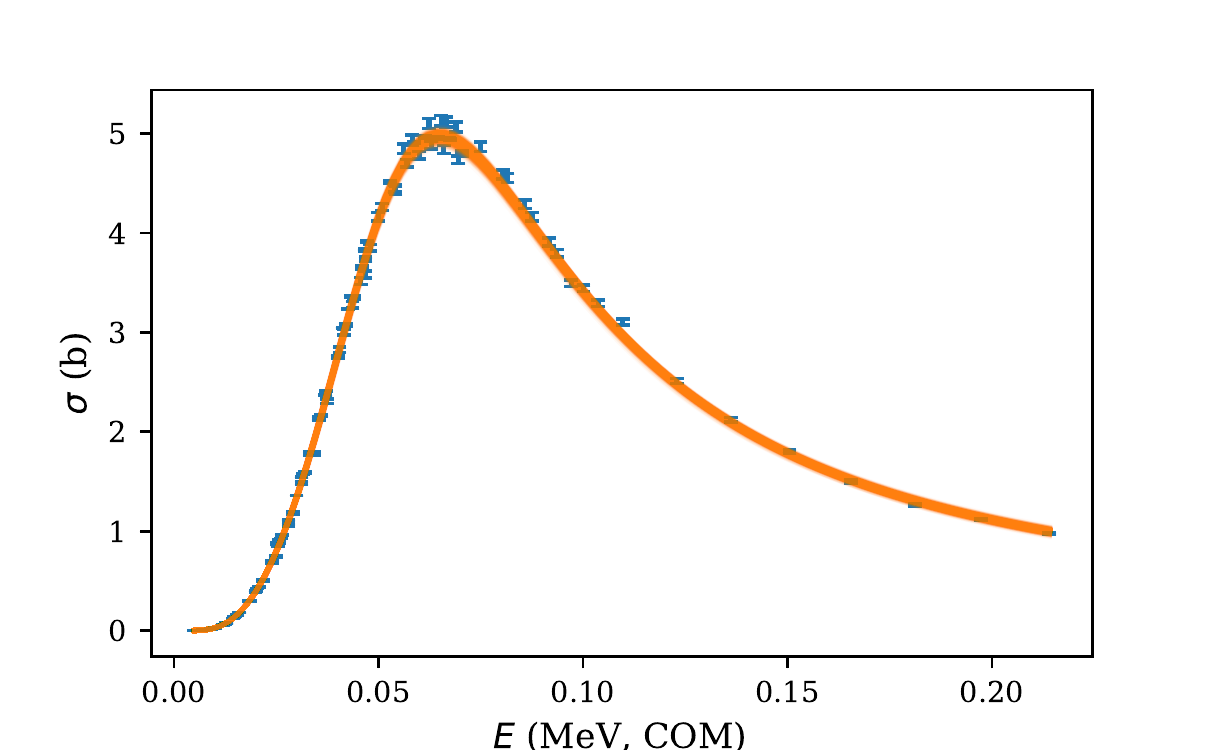}
    \caption{Cross section data (without extrinsic uncertainties) from the four sets in our final analysis compared to a series of curves generated (without normalization factors) from a subset of the chains for each of the $a_d$ values in the first column of Table~\ref{tab:modelAB_maxlnL}.}
    \label{fig:modelB_brush}
\end{figure}

As a visual indication of how well our parameter posteriors reproduce the data, a subset of the MCMC chain for $a_d=7.25$ fm and $a_n=3.633$ fm was used to generate several theory curves.
They are shown (without normalization factors) together with the experimental data (without extrinsic uncertainties) in Figure \ref{fig:modelB_brush}.

\begin{figure}
    \centering
    \includegraphics[width=\linewidth]{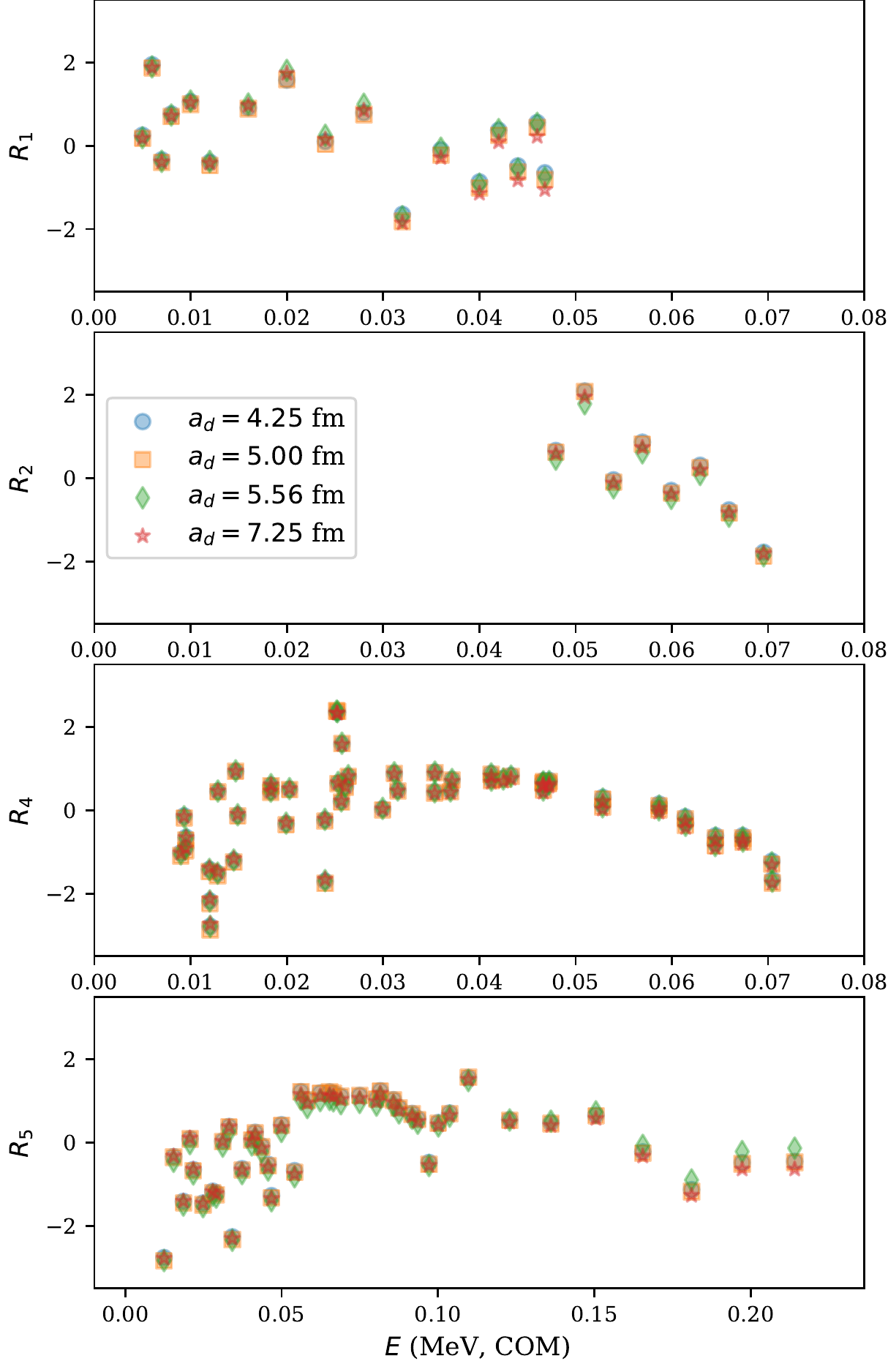}
    \caption{Residuals as defined by \eqref{eq:residuals} for the four data sets included in our final evaluation at four different values of $a_d$: 4.25 (blue circles), 5.00 (orange squares), 5.56 (green diamonds), and 7.25 fm (red stars). The residuals for different channel radii are so consistent that all four symbols often lie essentially on top of each other. As indicated in the figure, the top three panels share the same x-axis scale while the bottom panel extends over a wider energy range. Results were obtained using $R$-matrix Model B and statistical model $\mathcal{U}_{rf}$. $a_n$ was fixed at 3.633 fm. }
    \label{fig:modelB_data_residuals}
\end{figure}

The residuals, defined by Eq.~\eqref{eq:residuals}, for each of the four data sets used in our final analysis are shown in Figure \ref{fig:modelB_data_residuals}.
The residuals for the Jarmie, Arnold, and Conner data sets are very stable with respect to $a_d$ and across each data set's energy range: there is no observable, systematic trend.
The Brown residuals could be perceived to have a systematic decrease with energy, but there are only eight points in this data set and systematic behavior is more easily perceived in smaller data sets.
Each $a_d$ value has its own color and marker in the figure to make it easier to distinguish between them for the few points where they do not lie directly on top of each other.

\begin{figure}
    \centering
    \includegraphics{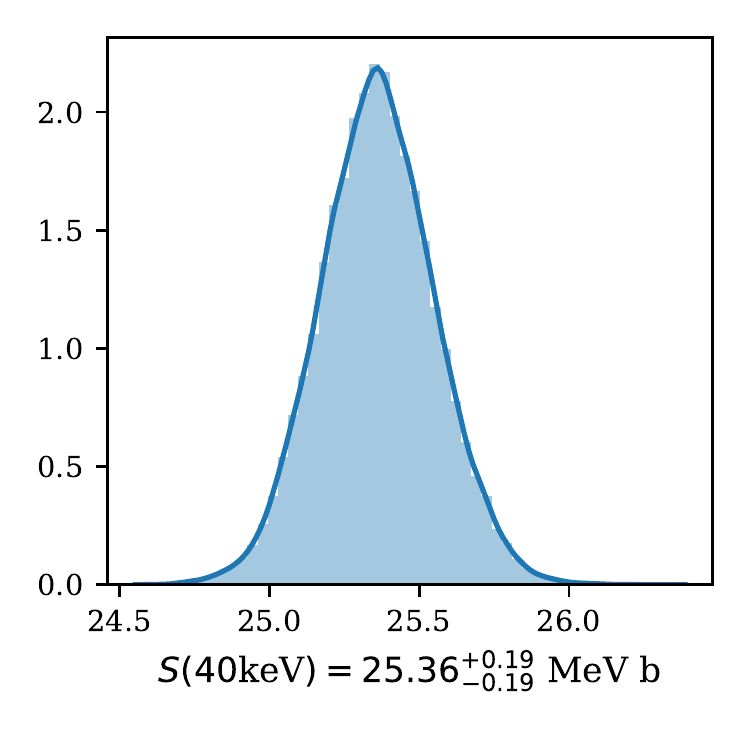}
    \caption{$S(40~\rm{keV})$ posterior generated with $R$-matrix Model B and statistical model $\mathcal{U}_{rf}$. Since results from all $a_d$ values in the first column of Table~\ref{tab:modelAB_maxlnL} are statistically consistent they are combined to produce one final result.}
    \label{fig:modelB_s40_bucket}
\end{figure}

\begin{figure}
    \centering
    \includegraphics{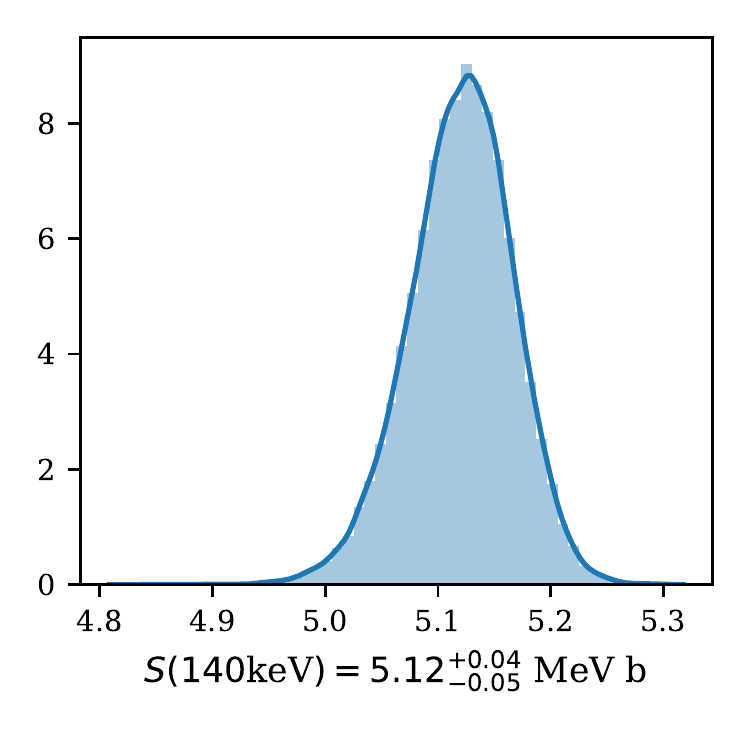}
    \caption{$S(140~\rm{keV})$ posterior generated with $R$-matrix Model B and statistical model $\mathcal{U}_{rf}$. Results from all $a_d$ values in the first column of Table~\ref{tab:modelAB_maxlnL} are combined, as in Fig.~\ref{fig:modelB_s40_bucket}.}
    \label{fig:modelB_s140_bucket}
\end{figure}

The Model B posterior for $S(40~\rm{keV})$ is presented in Figure \ref{fig:modelB_s40_bucket}.
Due to the consistency of both the $\ln{\mathcal{L}}$ values and physical observables, all $a_d$ values were binned together.
In this sense, the $S(40~\rm{keV})$ presented here is $a_d$- and $a_n$-independent.
The chains for the different $a_d$ configurations were not of equal length, so the shortest chain set the number of samples drawn from each. The result is $S(40~{\rm keV})=25.36 \pm 0.19~{\rm MeV}~{\rm b}$. It differs from the most recent evaluation by de Souza {\it et al.} in Ref.~\cite{deSouza:2019pmr} in two ways: the peak value is lower, and the width of the distribution is approximately twice as wide.
This widening is in no way associated with the reduced data set.
Similar widths were observed when Kobzev {\em et al.} data were included.
The $S(140~\rm{keV})$ posterior is shown in Figure \ref{fig:modelB_s140_bucket}.
Contributions from the different $a_d$ configurations were combined there in the same way as in Fig.~\ref{fig:modelB_s40_bucket}.

Finally, purely to ensure that our results are readily reproduced, we give the $R$-matrix and statistical parameters for $a_d=7.25$ fm and $a_n=3.633$ fm at $\max{\ln{\mathcal{L}}}_B$ in Tables \ref{tab:maxlnL_rpar} and \ref{tab:maxlnL_stat}.
We emphasize that $\theta_*^{(\mathcal{L})}$ is not representative of our analysis, which yields full posteriors. We provide it here purely as a benchmark.
\begin{table}
    \centering
    \begin{tabular}{|c|c|c|c|c|c|c|}
        $E_r$  & $\Gamma_{1d}$ & $\Gamma_{1n}$  & $\Gamma_{2d}$  & $\Gamma_{2n}$  & $U_e$  & $A_{1/2}$ \\
        MeV & MeV & MeV & MeV & MeV & eV & --- \\
        \hline
        0.071 & 0.046 & 0.075 & -11.964 & 1.598 & 5.923 & 0.008
    \end{tabular}
    \caption{$R$-matrix parameters at $\theta_{*}^{(\mathcal{L})}$ where $a_d=7.25~{\rm fm}$.} 
    \label{tab:maxlnL_rpar}
\end{table}
\begin{table}
    \centering
    \begin{tabular}{|c|c|c|c|c|c|c|c|}
        $\alpha_1$ & $\alpha_2$ & $\alpha_3$ & $\alpha_5$ & $f_1$ & $f_1$ & $f_3$ & $f_5$ \\
        \hline
        0.00 & 0.01 & 0.02 & 0.03 & 1.02 & 1.00 & 1.03 & 1.01
    \end{tabular}
    \caption{Statistical parameters at $\theta_{*}^{(\mathcal{L})}$ where $a_d=7.25~{\rm fm}$.} 
    \label{tab:maxlnL_stat}
\end{table}

\section{Conclusions}

\label{sec:conclusion}

We have demonstrated that the combination of $R$-matrix and Bayesian methods is a powerful tool for statistical inference in low-energy fusion reactions. 

In particular, we showed that:
\begin{itemize}
        \item Some low-energy $dt$ fusion data sets have underestimated or unreported point-to-point errors. Adding an extrinsic point-to-point error to each data set, as originally proposed in \textcite{deSouza:2019pmr}, makes it possible to obtain a statistically consistent description that incorporates four different data sets: Arnold, Brown, Conner, and Jarmie.
        
        \item Once the point-to-point errors are increased the common-mode errors becomes less certain, i.e., the widths obtained when they are inferred from data are larger. This leads to more uncertainty in the overall evaluation of $S(40~{\rm keV})$ than if extrinsic errors were not included.
                
        \item Our final result for $S(40~{\rm keV})$ is $25.36 \pm 0.19$ MeV b. This is consistent with \textcite{deSouza:2019pmr} result of $25.438^{+0.080}_{-0.089}$ MeV b but has an error bar that is a factor of two larger. We attribute the smaller error bar of \textcite{deSouza:2019pmr} to the fact that their sampler did not explore the full posterior of their $R$-matrix-plus-statistical model. 
        
        \item The available data on the $dt$ cross section below 250 keV leaves very little room for contributions from partial waves other than $l_d=0$ and $J^\pi=3/2^+$.
        
        \item If a single partial wave contributes to the reaction then quantum-mechanical unitarity defines a maximum possible cross section at a given energy as discussed in subsection~\ref{sec:unitarity}. At energies near 80 keV, the Conner data saturates this bound within uncertainties. Our ``best fit'' $R$-matrix parameters yield a value for the combination of partial widths that determines the maximum one-level, two-channel cross section, $\frac{\Gamma_d \Gamma_n}{(\Gamma_d + \Gamma_n)^2}$, of 0.236. This is very close to the unitary limit ($\Gamma_d=\Gamma_n$) value of 0.25. The proximity of this system to the unitary limit results in posteriors that exhibit a strong correlation between the reduced partial widths $\gamma_d^2$ and $\gamma_n^2$.
    
        \item For $E_{\rm c.m.}$ between 120 and 160 keV the data set of \textcite{Kobzev1966} disagrees at the 2--3 $\sigma$ level with the only other data set that extends above 100 keV, that of \textcite{Con52}. We allowed the energies of individual points in the Kobzev et al. data set to float within the reported energy uncertainty and found a systematic trend in the difference between the reported and optimal energies. Given the disagreement with the Conner {\emph et al.} data, and the absence of any discussion of this kind of systematic effect in Ref.~\cite{Kobzev1966}, we choose to omit the \textcite{Kobzev1966} data from our analysis. This does mean that our results above the resonance peak rely on only one data set. A modern measurement of ${}^3{\rm H}(d,n){}^4{\rm He}$ in the energy region above the resonance peak, that had clearly stated systematic and statistical uncertainties, would provide a valuable check on our analysis. It could also illuminate the issue of the unitarity limit discussed in the previous bullet. 
        
        \item We obtain results that are essentially the same for channel radii ranging from $a_d=4.25$ to $7.25$ fm and $a_n=3.633$ fm to $7.5$ fm by including a background level in the $3/2^+$ channel. This background level is not needed for the optimal fit at $a_d=4.25$ fm, but at larger deuteron-channel radii its parameters can be chosen to cancel the effects induced in $S(E)$ due to the increase in $a_d$ from 4.25 to 7.25 fm. Results do not vary with $a_n$ even if only a single channel is included. We note that results are not as stable with $a_d$ if the Kobzev {\emph et al.} data are included in the analysis.
        
        \item We find that the Brune parameterization~\cite{Brune2002} makes Monte Carlo sampling for the two-level model more efficient, because this parametrization largely decouples the two levels in the range of the experimental data. In contrast, if the Lane and Thomas parametrization~\cite{LaneThomas1958} is used, changes in the background level effectively alter the energy and reduced width parameters of the low-energy resonance. These correlations between the parameters associated with the two levels in the Lane and Thomas parameterization can make sampling slower to converge. 
\end{itemize}

It is generally the case that incremental improvements in an $R$-matrix analysis can be achieved by expanding the scope of data that are fitted. In the present situation, this could mean
\begin{enumerate}
    \item including $n+{}^4{\rm He}$ total cross section data around the $3/2^+$ resonance, as done by \textcite{Barker97};
    \item accounting for the very small anisotropy in the differential cross section~\cite{Bem97};
    \item extending the analysis to higher energies.
    \end{enumerate}
    The second and third options could provide a better estimate of the contribution of higher partial waves. However, as discussed in Sec.~\ref{sec:Rmatrix}, the energy range and partial waves considered here are well justified, and any improvement in the results presented here from expanding the scope would be incremental.

The points discussed so far in this summary are rather specific to the $dt$ fusion reaction. Since this reaction is dominated by a single, broad resonance it has features that are not shared by many other reactions to which $R$-matrix analysis is applied. What, then, are some general lessons that can be drawn from this study regarding Bayesian inference in $R$-matrix analyses?

First, our results emphasize the need to carefully explore correlations and ensure that the sampler has fully explored the posterior. Extended non-linear correlations between $R$-matrix parameters make canonical MCMC methods such as Metropolis-Hastings slow to converge---even in an ensemble sampler implementaton. Writing down an $R$-matrix formula and sampling all possible parameters may not be the best strategy. It is preferable to employ a parameterization that does not have (non-linear) correlations between different levels in the same channel, such as the Brune parameterization. And it is certainly not useful to sample parameters that have no impact on observables, such as the boundary condition parameter $B$. 

Second, not sampling the channel radius is beneficial at a practical level, since it eliminates one source of extended, non-linear correlations. Channel radius sampling is also deprecated for reasons of principle, since $R$-matrix results would be independent of channel radius if enough levels were included in each channel. For these reasons it is good to work on a grid of channel radii and demonstrate that the results for physical parameters and observables are similar across the grid. To achieve that similarity background levels will likely be necessary. Including them ensures a physics model that is sufficiently flexible to accommodate all data, not just at the resonance.

 Third, data uncertainties may need to be expanded through an ``extrinsic error'' formalism like the one that was used in Ref.~\cite{deSouza:2019pmr} which we incorporated in our analysis and modified here. Even a cursory examination of the point-to-point errors quoted for the Arnold data set~\cite{Arn53} makes it clear that an expansion of the error bars is required. For comparison we point out that if the Arnold, Brown, Conner, and Jarmie data sets were fed to AZURE2~\cite{Azuma2010} with the stated point-to-point errors a $\chi^2$ of approximately 25,700 for 166 data points would result. Such an analysis also leads to an $S(40~{\rm keV})$ of $28.08\pm 0.03$ MeV b, a spuriously precise value since the model is clearly incredibly unlikely to be correct.
 
 Bayesian methods provide several advantages in $R$-matrix modeling of nuclear reactions. They make it straightforward to supplement the $R$-matrix model by a ``statistical'' model that incorporates  known imperfections in the experiment in the analysis. They also provide access to the entire parameter posterior, and not just the region around the optimum $R$-matrix parameters. And the ability to specify priors on $R$-matrix and statistical-model parameters means that small effects can be tested in the analysis without destabilizing the parameter estimation. In the future we plan to exploit these benefits in other contexts, including ${}^3{\rm He}$-$\alpha$ elastic scattering and ${}^3{\rm He}(\alpha,\gamma){}^7{\rm Be}$.

\begin{acknowledgments}

We thank Christian Iliadis and Rafael deSouza for helpful discussions and their transparent communication of unpublished results. We acknowledge useful discussions with Dick Furnstahl, Richard Longland, and Sarah Wesolowski. We are grateful to James deBoer and Christian Iliadis for their careful reading and useful comments on the manuscript. This work was supported by the U.S. Department of Energy, National Nuclear Security Agency, under Award DE-NA0003883.

\end{acknowledgments}

\appendix

\section{Sample Details}
\label{sec:chain_details}

Here we present the corner plots of the different parameters sampled in Model B.
We grouped parameters ``like with like''. There are not noticeable correlations 
between $R$-matrix parameters, $\alpha_j$, and $f_j$: in this sense the correlations are block diagonal. 

\begin{figure}
    \centering
    \includegraphics[width=\linewidth]{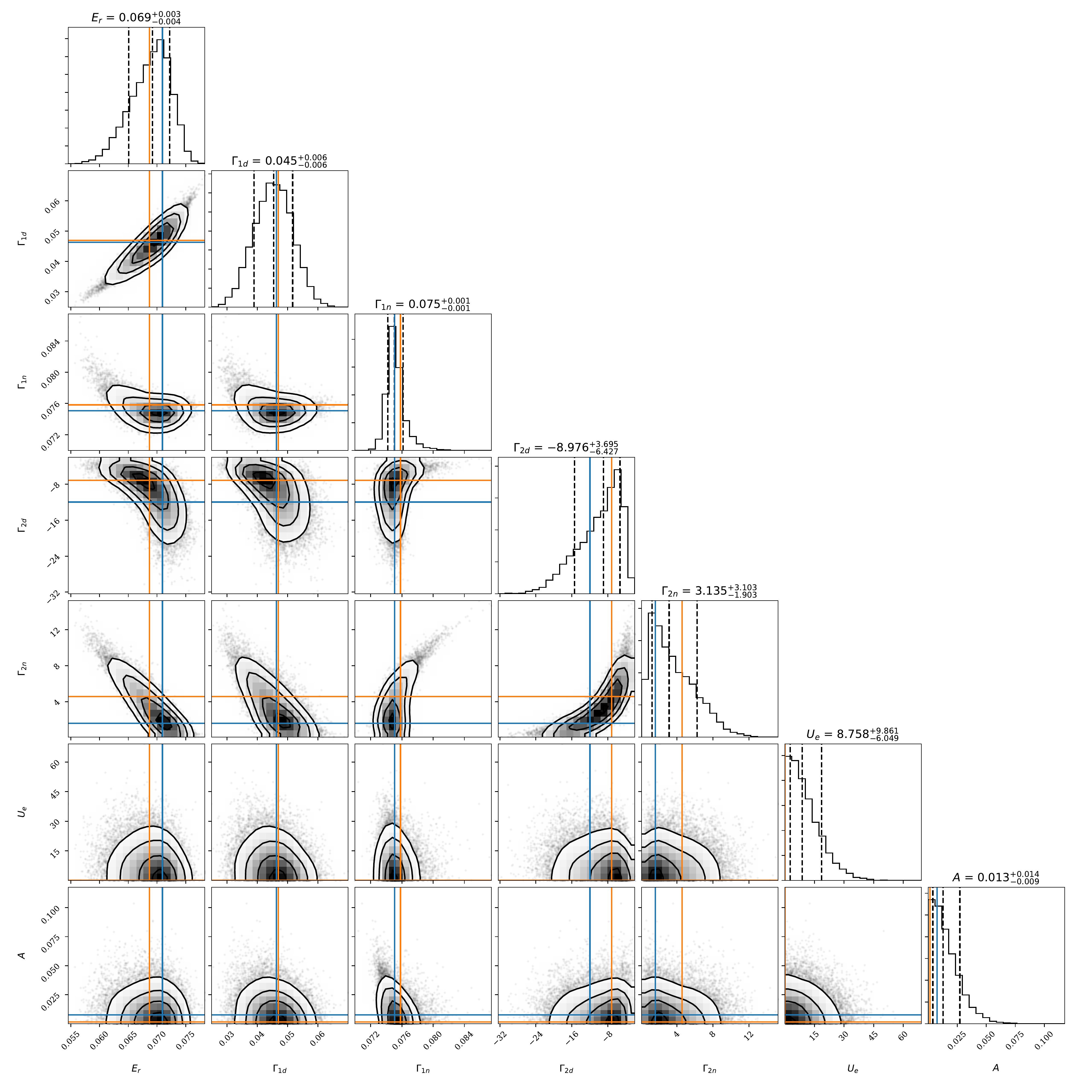}
    \caption{$R$-matrix parameter posteriors for $a_d=7.25$ fm and $a_n=3.633$ fm using $R$-matrix Model B and statistical model $\mathcal{U}_{rf}$}.
    \label{fig:blip_model_rpar_corner}
\end{figure}

\begin{figure}
    \centering
    \includegraphics[width=\linewidth]{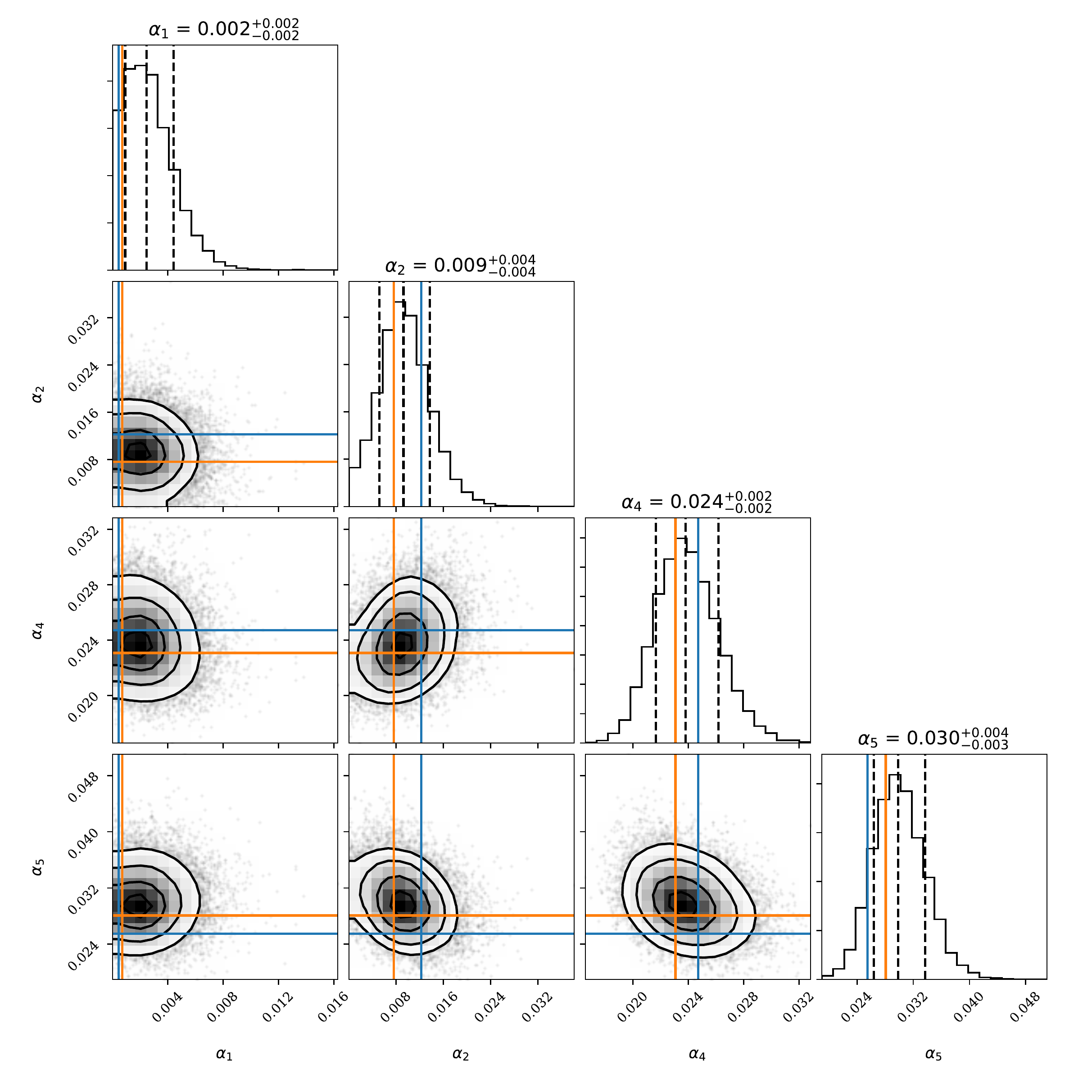}
    \caption{Relative extrinsic uncertainty parameter posteriors for $a_d=7.25$ fm and $a_n=3.633$ fm using $R$-matrix Model B and statistical model $\mathcal{U}_{rf}$.}
    \label{fig:blip_model_alpha_corner}
\end{figure}

\begin{figure}
    \centering
    \includegraphics[width=\linewidth]{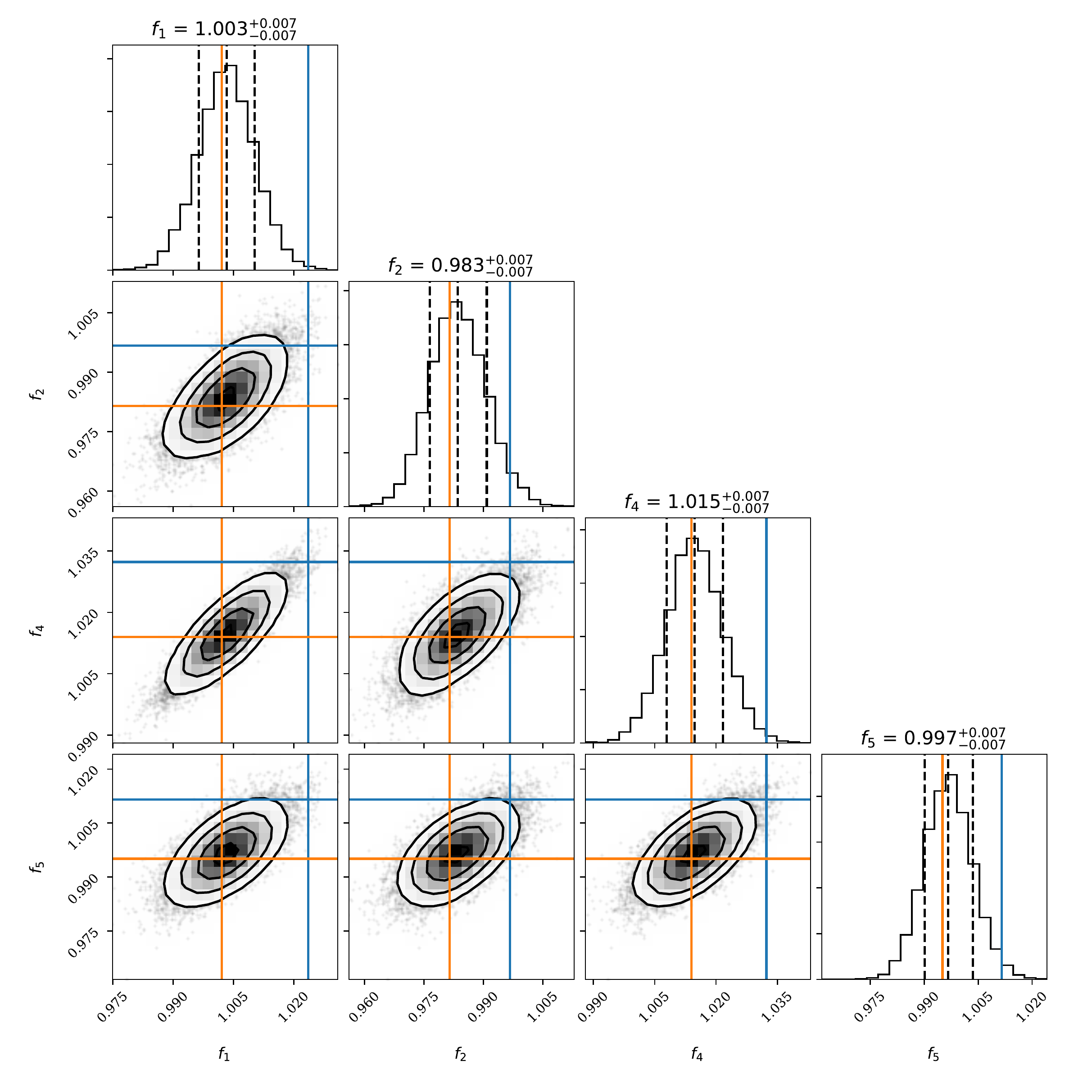}
    \caption{Normalization factor parameter posteriors for $a_d=7.25$ fm and $a_n=3.633$ fm using $R$-matrix Model B and statistical model $\mathcal{U}_{rf}$. These normalization factors were applied to the theory predictions.}
    \label{fig:blip_model_f_corner}
\end{figure}

Figures \ref{fig:blip_model_rpar_corner}, \ref{fig:blip_model_alpha_corner}, and \ref{fig:blip_model_f_corner} display the one- and two-dimensional posteriors of the $R$-matrix parameters, extrinsic uncertainties, and normalization factors respectively for the choice $a_d=7.25$ fm and $a_n=3.633$ fm.
In all three figures, the blue lines represent the $\theta_{*}^{(\mathcal{L})}$ values and the orange lines represent the $\theta_{*}^{(\mathcal{P})}$ values.

As can be seen in Fig.~\ref{fig:blip_model_f_corner}, the median values of the normalization factors match up fairly well with \cite{deSouza:2019pmr}.
However, our results consistently return distributions that are approximately twice as wide.
This increase in normalization factor uncertainty, obtained exclusively when extrinsic uncertainties are simultaneously sampled, directly increases the width of $S(40~\rm{keV})$.

\section{Sampler diagnostics}
\label{sec:sampler_diagnostics}

Chain convergence is based on integrated autocorrelation time, $\tau$, as provided by the {\tt emcee} package, documented \href{https://emcee.readthedocs.io/en/stable/user/autocorr/}{here}.
Once we obtained a reliable estimate of $\tau$ for each parameter, we thinned all of the chains by the maximum $\tau$.
Typical $\tau_{\max}$ values were 3000 to 6000 in our final analysis using $R$-matrix Model B.
To obtain these numbers
each run was pre-thinned by a factor of approximately 20.
The $\tau_{\max}$ for the pre-thinned chain was then computed to be 150 to 300, with the final estimates obtained after multiplying by the pre-thinning factor.

Autocorrelation times were calculated \textit{after} the burn-in period had been discarded.
In some cases, walkers in the ensemble never found the maximum region of the posterior.
Those walkers were identified by the mean of their $\ln\mathcal{P}$ values.
This was a small percentage of the ensemble, never more than 10\%, so discarding them from the final evaluation was not only justified by their relatively low $\ln{\mathcal{P}}$ values but also fairly insignificant.
Their primary contribution was disproportionately affecting the calculation of $\tau$.

Thinned chains were further analyzed to confirm that the mean of each chain was stable, ensuring equilibrium had been reached.

\section{Alternative Sampling Methods}
\label{sec:alt_methods}

When the channel radii were treated as parameters, meaning that they were sampled, we found parallel tempering (via \texttt{ptemcee} \cite{Vousden_2015,Foreman_Mackey_2013}) to be useful.
The technique tempers difficult-to-sample likelihoods according to
\begin{equation}
    \label{eq:parallel_tempering}
    p(\mathcal{D}|\theta,\mathcal{I})^{1/T}~,
\end{equation}
where $T$ is a so-called temperature, allowing several chains to be run at different temperatures simultaneously.
The benefit of the algorithm comes with the exchange of proposals \textit{between} temperature chains.
Higher temperature chains move through less complex multidimensional posteriors, only exploring the most prominent features.
The chain corresponding to $T=1$ is the target distribution, but by exchanging samples with higher-temperature chains walkers gain the ability to sample the posterior on larger distance scales than would otherwise be possible. Consequently the $T=1$ chain is able to obtain a better, more accurate representation of posteriors that contain complex, non-linear structures. For our final results, where channel radii were fixed, we did not find parallel tempering to be necessary, but it did provide significant insight into the non-linear, extended correlations between the parameters.

Instead or in addition to better sampling techniques, one can also investigate different parameterizations.
We explored two such alternative parameterizations.
One, we generated an orthogonal transformation by performing a principal component analysis (PCA) on a subset of our typical $R$-matrix-parameter chains.
This generates a linear, orthogonal transformation which we can use to sample uncorrelated parameters.
Of course, in the case where nonlinear correlations were dominant, this linear transformation was ineffective and offered little to no improvement in sampling efficiency.

The other alternative parameterization we attempted to sample was based on a nonlinear transformation motivated by the $R$-matrix argument presented in Section~\ref{sec:conclusion}.
This alternative parameter set can be employed to de-correlate the pairs $\{\gamma_d^2,a_d\}$ and $\{\gamma_n^2,a_n\}$.
Using \eqref{eq:ag2_corr} and the values for $\alpha_c$ fit to the observed correlations in Figure \ref{fig:gc2_ac_correlation}, we sampled the product of $\gamma_c^2a_c$ and its orthogonal construction $\gamma_c^4 - \frac{1}{\alpha_c}a_c^2$.
Again, our final results are based on models where the channel radii were fixed, so this method, while demonstrating noticeable improvement to sampling efficiency, specifically in terms of autocorrelation times, was ultimately not necessary.


%


\begin{thebibliography}{29}%
\makeatletter
\providecommand \@ifxundefined [1]{%
 \@ifx{#1\undefined}
}%
\providecommand \@ifnum [1]{%
 \ifnum #1\expandafter \@firstoftwo
 \else \expandafter \@secondoftwo
 \fi
}%
\providecommand \@ifx [1]{%
 \ifx #1\expandafter \@firstoftwo
 \else \expandafter \@secondoftwo
 \fi
}%
\providecommand \natexlab [1]{#1}%
\providecommand \enquote  [1]{``#1''}%
\providecommand \bibnamefont  [1]{#1}%
\providecommand \bibfnamefont [1]{#1}%
\providecommand \citenamefont [1]{#1}%
\providecommand \href@noop [0]{\@secondoftwo}%
\providecommand \href [0]{\begingroup \@sanitize@url \@href}%
\providecommand \@href[1]{\@@startlink{#1}\@@href}%
\providecommand \@@href[1]{\endgroup#1\@@endlink}%
\providecommand \@sanitize@url [0]{\catcode `\\12\catcode `\$12\catcode
  `\&12\catcode `\#12\catcode `\^12\catcode `\_12\catcode `\%12\relax}%
\providecommand \@@startlink[1]{}%
\providecommand \@@endlink[0]{}%
\providecommand \url  [0]{\begingroup\@sanitize@url \@url }%
\providecommand \@url [1]{\endgroup\@href {#1}{\urlprefix }}%
\providecommand \urlprefix  [0]{URL }%
\providecommand \Eprint [0]{\href }%
\providecommand \doibase [0]{https://doi.org/}%
\providecommand \selectlanguage [0]{\@gobble}%
\providecommand \bibinfo  [0]{\@secondoftwo}%
\providecommand \bibfield  [0]{\@secondoftwo}%
\providecommand \translation [1]{[#1]}%
\providecommand \BibitemOpen [0]{}%
\providecommand \bibitemStop [0]{}%
\providecommand \bibitemNoStop [0]{.\EOS\space}%
\providecommand \EOS [0]{\spacefactor3000\relax}%
\providecommand \BibitemShut  [1]{\csname bibitem#1\endcsname}%
\let\auto@bib@innerbib\@empty
\bibitem [{\citenamefont {Conner}\ \emph {et~al.}(1952)\citenamefont {Conner},
  \citenamefont {Bonner},\ and\ \citenamefont {Smith}}]{Con52}%
  \BibitemOpen
  \bibfield  {author} {\bibinfo {author} {\bibfnamefont {J.~P.}\ \bibnamefont
  {Conner}}, \bibinfo {author} {\bibfnamefont {T.~W.}\ \bibnamefont {Bonner}},\
  and\ \bibinfo {author} {\bibfnamefont {J.~R.}\ \bibnamefont {Smith}},\
  }\bibfield  {title} {\bibinfo {title} {A study of the
  $\mathrm{H}^3(d,n)\mathrm{He}^{4}$ reaction},\ }\href
  {https://doi.org/10.1103/PhysRev.88.468} {\bibfield  {journal} {\bibinfo
  {journal} {Phys. Rev.}\ }\textbf {\bibinfo {volume} {88}},\ \bibinfo {pages}
  {468} (\bibinfo {year} {1952})}\BibitemShut {NoStop}%
\bibitem [{\citenamefont {Arnold}\ \emph {et~al.}(1953)\citenamefont {Arnold},
  \citenamefont {Phillips}, \citenamefont {Sawyer}, \citenamefont {Stovall},\
  and\ \citenamefont {Tuck}}]{Arn53}%
  \BibitemOpen
  \bibfield  {author} {\bibinfo {author} {\bibfnamefont {W.~R.}\ \bibnamefont
  {Arnold}}, \bibinfo {author} {\bibfnamefont {J.~A.}\ \bibnamefont
  {Phillips}}, \bibinfo {author} {\bibfnamefont {G.~A.}\ \bibnamefont
  {Sawyer}}, \bibinfo {author} {\bibfnamefont {E.~J.}\ \bibnamefont {Stovall},
  \bibfnamefont {Jr.}},\ and\ \bibinfo {author} {\bibfnamefont {J.~L.}\
  \bibnamefont {Tuck}},\ }\href
  {https://permalink.lanl.gov/object/tr?what=info:lanl-repo/lareport/LA-01479}
  {\emph {\bibinfo {title} {Absolute cross section for the reaction
  $\mathrm{T(d,n)}^4\mathrm{He}$ from 10 to 120 keV}}},\ \bibinfo {type} {Tech.
  Rep.}\ \bibinfo {number} {LA-1479}\ (\bibinfo  {institution} {Los Alamos
  Scientific Laboratory},\ \bibinfo {year} {1953})\BibitemShut {NoStop}%
\bibitem [{\citenamefont {Kobzev}\ \emph {et~al.}(1966)\citenamefont {Kobzev},
  \citenamefont {Salatskij},\ and\ \citenamefont {Telezhnikov}}]{Kobzev1966}%
  \BibitemOpen
  \bibfield  {author} {\bibinfo {author} {\bibfnamefont {A.}~\bibnamefont
  {Kobzev}}, \bibinfo {author} {\bibfnamefont {V.}~\bibnamefont {Salatskij}},\
  and\ \bibinfo {author} {\bibfnamefont {S.}~\bibnamefont {Telezhnikov}},\
  }\bibfield  {title} {\bibinfo {title} {Differential cross sections for the
  reaction {D}$(t,\alpha)n$ at 115-1650 ke{V}},\ }\href@noop {} {\bibfield
  {journal} {\bibinfo  {journal} {Sov. J. Nucl. Phys}\ }\textbf {\bibinfo
  {volume} {3}},\ \bibinfo {pages} {774} (\bibinfo {year} {1966})}\BibitemShut
  {NoStop}%
\bibitem [{\citenamefont {Jarmie}\ \emph {et~al.}(1984)\citenamefont {Jarmie},
  \citenamefont {Brown},\ and\ \citenamefont {Hardekopf}}]{Jar84}%
  \BibitemOpen
  \bibfield  {author} {\bibinfo {author} {\bibfnamefont {N.}~\bibnamefont
  {Jarmie}}, \bibinfo {author} {\bibfnamefont {R.~E.}\ \bibnamefont {Brown}},\
  and\ \bibinfo {author} {\bibfnamefont {R.~A.}\ \bibnamefont {Hardekopf}},\
  }\bibfield  {title} {\bibinfo {title} {Fusion-energy reaction
  $^{2}\mathrm{H}(\mathrm{t},\ensuremath{\alpha})\mathrm{n}$ from
  ${E}_{\mathrm{t}}=12.5$ to 117 ke{V}},\ }\href
  {https://doi.org/10.1103/PhysRevC.29.2031} {\bibfield  {journal} {\bibinfo
  {journal} {Phys. Rev. C}\ }\textbf {\bibinfo {volume} {29}},\ \bibinfo
  {pages} {2031} (\bibinfo {year} {1984})}\BibitemShut {NoStop}%
\bibitem [{\citenamefont {Brown}\ \emph {et~al.}(1987)\citenamefont {Brown},
  \citenamefont {Jarmie},\ and\ \citenamefont {Hale}}]{Bro87}%
  \BibitemOpen
  \bibfield  {author} {\bibinfo {author} {\bibfnamefont {R.~E.}\ \bibnamefont
  {Brown}}, \bibinfo {author} {\bibfnamefont {N.}~\bibnamefont {Jarmie}},\ and\
  \bibinfo {author} {\bibfnamefont {G.~M.}\ \bibnamefont {Hale}},\ }\bibfield
  {title} {\bibinfo {title} {Fusion-energy reaction
  $^{3}${H}(d,\ensuremath{\alpha})n at low energies},\ }\href
  {https://doi.org/10.1103/PhysRevC.35.1999} {\bibfield  {journal} {\bibinfo
  {journal} {Phys. Rev. C}\ }\textbf {\bibinfo {volume} {35}},\ \bibinfo
  {pages} {1999} (\bibinfo {year} {1987})}\BibitemShut {NoStop}%
\bibitem [{\citenamefont {Nollett}\ and\ \citenamefont {Burles}(2000)}]{Nol00}%
  \BibitemOpen
  \bibfield  {author} {\bibinfo {author} {\bibfnamefont {K.~M.}\ \bibnamefont
  {Nollett}}\ and\ \bibinfo {author} {\bibfnamefont {S.}~\bibnamefont
  {Burles}},\ }\bibfield  {title} {\bibinfo {title} {Estimating reaction rates
  and uncertainties for primordial nucleosynthesis},\ }\href
  {https://doi.org/10.1103/PhysRevD.61.123505} {\bibfield  {journal} {\bibinfo
  {journal} {Phys. Rev. D}\ }\textbf {\bibinfo {volume} {61}},\ \bibinfo
  {pages} {123505} (\bibinfo {year} {2000})}\BibitemShut {NoStop}%
\bibitem [{\citenamefont {Hupin}\ \emph {et~al.}(2019)\citenamefont {Hupin},
  \citenamefont {Quaglioni},\ and\ \citenamefont {Navr\'atil}}]{Hupin:2018biv}%
  \BibitemOpen
  \bibfield  {author} {\bibinfo {author} {\bibfnamefont {G.}~\bibnamefont
  {Hupin}}, \bibinfo {author} {\bibfnamefont {S.}~\bibnamefont {Quaglioni}},\
  and\ \bibinfo {author} {\bibfnamefont {P.}~\bibnamefont {Navr\'atil}},\
  }\bibfield  {title} {\bibinfo {title} {{Ab initio predictions for polarized
  deuterium-tritium thermonuclear fusion}},\ }\href
  {https://doi.org/10.1038/s41467-018-08052-6} {\bibfield  {journal} {\bibinfo
  {journal} {Nature Commun.}\ }\textbf {\bibinfo {volume} {10}},\ \bibinfo
  {pages} {351} (\bibinfo {year} {2019})},\ \Eprint
  {https://arxiv.org/abs/1803.11378} {arXiv:1803.11378 [nucl-th]} \BibitemShut
  {NoStop}%
\bibitem [{\citenamefont {Brown}\ and\ \citenamefont
  {Hale}(2014)}]{Brown:2013zla}%
  \BibitemOpen
  \bibfield  {author} {\bibinfo {author} {\bibfnamefont {L.~S.}\ \bibnamefont
  {Brown}}\ and\ \bibinfo {author} {\bibfnamefont {G.~M.}\ \bibnamefont
  {Hale}},\ }\bibfield  {title} {\bibinfo {title} {{Field Theory of the $d + t
  \to n + \alpha$ Reaction Dominated by a $^5$He$^*$ Unstable Particle}},\
  }\href {https://doi.org/10.1103/PhysRevC.89.014622} {\bibfield  {journal}
  {\bibinfo  {journal} {Phys. Rev. C}\ }\textbf {\bibinfo {volume} {89}},\
  \bibinfo {pages} {014622} (\bibinfo {year} {2014})},\ \Eprint
  {https://arxiv.org/abs/1308.0347} {arXiv:1308.0347 [nucl-th]} \BibitemShut
  {NoStop}%
\bibitem [{\citenamefont {Hale}\ \emph {et~al.}(1987)\citenamefont {Hale},
  \citenamefont {Brown},\ and\ \citenamefont {Jarmie}}]{Hale:1987}%
  \BibitemOpen
  \bibfield  {author} {\bibinfo {author} {\bibfnamefont {G.~M.}\ \bibnamefont
  {Hale}}, \bibinfo {author} {\bibfnamefont {R.~E.}\ \bibnamefont {Brown}},\
  and\ \bibinfo {author} {\bibfnamefont {N.}~\bibnamefont {Jarmie}},\
  }\bibfield  {title} {\bibinfo {title} {Pole structure of the
  ${J}^{\mathrm{\ensuremath{\pi}}}$${=3/2}^{+}$ resonance in
  $^{5}\mathrm{He}$},\ }\href {https://doi.org/10.1103/PhysRevLett.59.763}
  {\bibfield  {journal} {\bibinfo  {journal} {Phys. Rev. Lett.}\ }\textbf
  {\bibinfo {volume} {59}},\ \bibinfo {pages} {763} (\bibinfo {year}
  {1987})}\BibitemShut {NoStop}%
\bibitem [{\citenamefont {Bosch}\ and\ \citenamefont
  {Hale}(1992)}]{Bosch:1992}%
  \BibitemOpen
  \bibfield  {author} {\bibinfo {author} {\bibfnamefont {H.-S.}\ \bibnamefont
  {Bosch}}\ and\ \bibinfo {author} {\bibfnamefont {G.}~\bibnamefont {Hale}},\
  }\bibfield  {title} {\bibinfo {title} {Improved formulas for fusion
  cross-sections and thermal reactivities},\ }\href
  {https://doi.org/10.1088/0029-5515/32/4/i07} {\bibfield  {journal} {\bibinfo
  {journal} {Nuclear Fusion}\ }\textbf {\bibinfo {volume} {32}},\ \bibinfo
  {pages} {611} (\bibinfo {year} {1992})}\BibitemShut {NoStop}%
\bibitem [{\citenamefont {Bosch}\ and\ \citenamefont
  {Hale}(1993)}]{Bosch:1993}%
  \BibitemOpen
  \bibfield  {author} {\bibinfo {author} {\bibfnamefont {H.-S.}\ \bibnamefont
  {Bosch}}\ and\ \bibinfo {author} {\bibfnamefont {G.}~\bibnamefont {Hale}},\
  }\bibfield  {title} {\bibinfo {title} {Improved formulas for fusion
  cross-sections and thermal reactivities, {E}rratum},\ }\href
  {https://doi.org/10.1088/0029-5515/33/12/513} {\bibfield  {journal} {\bibinfo
   {journal} {Nuclear Fusion}\ }\textbf {\bibinfo {volume} {33}},\ \bibinfo
  {pages} {1919} (\bibinfo {year} {1993})}\BibitemShut {NoStop}%
\bibitem [{\citenamefont {de~Souza}\ \emph {et~al.}(2019)\citenamefont
  {de~Souza}, \citenamefont {Boston}, \citenamefont {Coc},\ and\ \citenamefont
  {Iliadis}}]{deSouza:2019pmr}%
  \BibitemOpen
  \bibfield  {author} {\bibinfo {author} {\bibfnamefont {R.~S.}\ \bibnamefont
  {de~Souza}}, \bibinfo {author} {\bibfnamefont {S.~R.}\ \bibnamefont
  {Boston}}, \bibinfo {author} {\bibfnamefont {A.}~\bibnamefont {Coc}},\ and\
  \bibinfo {author} {\bibfnamefont {C.}~\bibnamefont {Iliadis}},\ }\bibfield
  {title} {\bibinfo {title} {{Thermonuclear fusion rates for tritium +
  deuterium using Bayesian methods}},\ }\href
  {https://doi.org/10.1103/PhysRevC.99.014619, 10.1103/PhysRevC.00.004600}
  {\bibfield  {journal} {\bibinfo  {journal} {Phys. Rev.}\ }\textbf {\bibinfo
  {volume} {C99}},\ \bibinfo {pages} {014619} (\bibinfo {year} {2019})},\
  \Eprint {https://arxiv.org/abs/1901.04857} {arXiv:1901.04857 [nucl-th]}
  \BibitemShut {NoStop}%
\bibitem [{\citenamefont {Lane}\ and\ \citenamefont
  {Thomas}(1958)}]{LaneThomas1958}%
  \BibitemOpen
  \bibfield  {author} {\bibinfo {author} {\bibfnamefont {A.~M.}\ \bibnamefont
  {Lane}}\ and\ \bibinfo {author} {\bibfnamefont {R.~G.}\ \bibnamefont
  {Thomas}},\ }\bibfield  {title} {\bibinfo {title} {R-matrix theory of nuclear
  reactions},\ }\href {https://doi.org/10.1103/RevModPhys.30.257} {\bibfield
  {journal} {\bibinfo  {journal} {Rev. Mod. Phys.}\ }\textbf {\bibinfo {volume}
  {30}},\ \bibinfo {pages} {257} (\bibinfo {year} {1958})}\BibitemShut
  {NoStop}%
\bibitem [{\citenamefont {Wigner}\ and\ \citenamefont
  {Eisenbud}(1947)}]{Wig47}%
  \BibitemOpen
  \bibfield  {author} {\bibinfo {author} {\bibfnamefont {E.~P.}\ \bibnamefont
  {Wigner}}\ and\ \bibinfo {author} {\bibfnamefont {L.}~\bibnamefont
  {Eisenbud}},\ }\bibfield  {title} {\bibinfo {title} {Higher angular momenta
  and long range interaction in resonance reactions},\ }\href
  {https://doi.org/10.1103/PhysRev.72.29} {\bibfield  {journal} {\bibinfo
  {journal} {Phys. Rev.}\ }\textbf {\bibinfo {volume} {72}},\ \bibinfo {pages}
  {29} (\bibinfo {year} {1947})}\BibitemShut {NoStop}%
\bibitem [{\citenamefont {Brune}(2002)}]{Brune2002}%
  \BibitemOpen
  \bibfield  {author} {\bibinfo {author} {\bibfnamefont {C.~R.}\ \bibnamefont
  {Brune}},\ }\bibfield  {title} {\bibinfo {title} {Alternative parametrization
  of \textit{R}-matrix theory},\ }\href
  {https://doi.org/10.1103/PhysRevC.66.044611} {\bibfield  {journal} {\bibinfo
  {journal} {Phys. Rev. C}\ }\textbf {\bibinfo {volume} {66}},\ \bibinfo
  {pages} {044611} (\bibinfo {year} {2002})}\BibitemShut {NoStop}%
\bibitem [{\citenamefont {Hale}\ \emph {et~al.}(2014)\citenamefont {Hale},
  \citenamefont {Brown},\ and\ \citenamefont {Paris}}]{Hale:2013ama}%
  \BibitemOpen
  \bibfield  {author} {\bibinfo {author} {\bibfnamefont {G.~M.}\ \bibnamefont
  {Hale}}, \bibinfo {author} {\bibfnamefont {L.~S.}\ \bibnamefont {Brown}},\
  and\ \bibinfo {author} {\bibfnamefont {M.~W.}\ \bibnamefont {Paris}},\
  }\bibfield  {title} {\bibinfo {title} {{Effective field theory as a limit of
  \textit{R}-matrix theory for light nuclear reactions}},\ }\href
  {https://doi.org/10.1103/PhysRevC.89.014623} {\bibfield  {journal} {\bibinfo
  {journal} {Phys. Rev.}\ }\textbf {\bibinfo {volume} {C89}},\ \bibinfo {pages}
  {014623} (\bibinfo {year} {2014})},\ \Eprint
  {https://arxiv.org/abs/1308.0348} {arXiv:1308.0348 [nucl-th]} \BibitemShut
  {NoStop}%
\bibitem [{\citenamefont {deBoer}\ \emph {et~al.}(2017)\citenamefont {deBoer},
  \citenamefont {G\"orres}, \citenamefont {Wiescher}, \citenamefont {Azuma},
  \citenamefont {Best}, \citenamefont {Brune}, \citenamefont {Fields},
  \citenamefont {Jones}, \citenamefont {Pignatari}, \citenamefont {Sayre},
  \citenamefont {Smith}, \citenamefont {Timmes},\ and\ \citenamefont
  {Uberseder}}]{deB17}%
  \BibitemOpen
  \bibfield  {author} {\bibinfo {author} {\bibfnamefont {R.~J.}\ \bibnamefont
  {deBoer}}, \bibinfo {author} {\bibfnamefont {J.}~\bibnamefont {G\"orres}},
  \bibinfo {author} {\bibfnamefont {M.}~\bibnamefont {Wiescher}}, \bibinfo
  {author} {\bibfnamefont {R.~E.}\ \bibnamefont {Azuma}}, \bibinfo {author}
  {\bibfnamefont {A.}~\bibnamefont {Best}}, \bibinfo {author} {\bibfnamefont
  {C.~R.}\ \bibnamefont {Brune}}, \bibinfo {author} {\bibfnamefont {C.~E.}\
  \bibnamefont {Fields}}, \bibinfo {author} {\bibfnamefont {S.}~\bibnamefont
  {Jones}}, \bibinfo {author} {\bibfnamefont {M.}~\bibnamefont {Pignatari}},
  \bibinfo {author} {\bibfnamefont {D.}~\bibnamefont {Sayre}}, \bibinfo
  {author} {\bibfnamefont {K.}~\bibnamefont {Smith}}, \bibinfo {author}
  {\bibfnamefont {F.~X.}\ \bibnamefont {Timmes}},\ and\ \bibinfo {author}
  {\bibfnamefont {E.}~\bibnamefont {Uberseder}},\ }\bibfield  {title} {\bibinfo
  {title} {The ${{}^{12}{\rm C}(\ensuremath{\alpha},
  \ensuremath{\gamma})^{16}{\rm O}}$ reaction and its implications for stellar
  helium burning},\ }\href {https://doi.org/10.1103/RevModPhys.89.035007}
  {\bibfield  {journal} {\bibinfo  {journal} {Rev. Mod. Phys.}\ }\textbf
  {\bibinfo {volume} {89}},\ \bibinfo {pages} {035007} (\bibinfo {year}
  {2017})}\BibitemShut {NoStop}%
\bibitem [{\citenamefont {B{\' e}m}\ \emph {et~al.}(1997)\citenamefont {B{\'
  e}m}, \citenamefont {Kroha}, \citenamefont {Mare{\v{s}}}, \citenamefont
  {{\v{S}}ime{\v{c}}kov{\' a}}, \citenamefont {Trgi{\v{n}}ov{\' a}},\ and\
  \citenamefont {Ver{\v{c}}im{\' a}k}}]{Bem97}%
  \BibitemOpen
  \bibfield  {author} {\bibinfo {author} {\bibfnamefont {P.}~\bibnamefont {B{\'
  e}m}}, \bibinfo {author} {\bibfnamefont {V.}~\bibnamefont {Kroha}}, \bibinfo
  {author} {\bibfnamefont {J.}~\bibnamefont {Mare{\v{s}}}}, \bibinfo {author}
  {\bibfnamefont {E.}~\bibnamefont {{\v{S}}ime{\v{c}}kov{\' a}}}, \bibinfo
  {author} {\bibfnamefont {M.}~\bibnamefont {Trgi{\v{n}}ov{\' a}}},\ and\
  \bibinfo {author} {\bibfnamefont {P.}~\bibnamefont {Ver{\v{c}}im{\' a}k}},\
  }\bibfield  {title} {\bibinfo {title} {Angular anisotropy of the
  ${}^3\mathrm{H}(d,\alpha)n$ reaction at deuteron energies below 200 {keV}},\
  }\href {https://doi.org/10.1007/s006010050055} {\bibfield  {journal}
  {\bibinfo  {journal} {Few-Body Systems}\ }\textbf {\bibinfo {volume} {22}},\
  \bibinfo {pages} {77} (\bibinfo {year} {1997})}\BibitemShut {NoStop}%
\bibitem [{\citenamefont {Mori}(1972)}]{Mor72}%
  \BibitemOpen
  \bibfield  {author} {\bibinfo {author} {\bibfnamefont {A.}~\bibnamefont
  {Mori}},\ }\bibfield  {title} {\bibinfo {title} {Boundary-condition constants
  of the {L}ane-{R}obson calculable theory},\ }\href
  {https://doi.org/10.1103/PhysRevC.5.1795} {\bibfield  {journal} {\bibinfo
  {journal} {Phys. Rev. C}\ }\textbf {\bibinfo {volume} {5}},\ \bibinfo {pages}
  {1795} (\bibinfo {year} {1972})}\BibitemShut {NoStop}%
\bibitem [{\citenamefont {{Barker}}(1972)}]{Barker72}%
  \BibitemOpen
  \bibfield  {author} {\bibinfo {author} {\bibfnamefont {F.~C.}\ \bibnamefont
  {{Barker}}},\ }\bibfield  {title} {\bibinfo {title} {{The boundary condition
  parameter in \textit{R}-matrix theory}},\ }\href
  {https://doi.org/10.1071/PH720341} {\bibfield  {journal} {\bibinfo  {journal}
  {Australian Journal of Physics}\ }\textbf {\bibinfo {volume} {25}},\ \bibinfo
  {pages} {341} (\bibinfo {year} {1972})}\BibitemShut {NoStop}%
\bibitem [{\citenamefont {Assenbaum}\ \emph {et~al.}(1987)\citenamefont
  {Assenbaum}, \citenamefont {Langanke},\ and\ \citenamefont {Rolfs}}]{Ass87}%
  \BibitemOpen
  \bibfield  {author} {\bibinfo {author} {\bibfnamefont {H.~J.}\ \bibnamefont
  {Assenbaum}}, \bibinfo {author} {\bibfnamefont {K.}~\bibnamefont
  {Langanke}},\ and\ \bibinfo {author} {\bibfnamefont {C.}~\bibnamefont
  {Rolfs}},\ }\bibfield  {title} {\bibinfo {title} {Effects of electron
  screening on low-energy fusion cross sections},\ }\href
  {https://doi.org/10.1007/BF01289572} {\bibfield  {journal} {\bibinfo
  {journal} {Zeitschrift f{\" u}r Physik A: Atomic Nuclei}\ }\textbf {\bibinfo
  {volume} {327}},\ \bibinfo {pages} {461} (\bibinfo {year}
  {1987})}\BibitemShut {NoStop}%
\bibitem [{\citenamefont {Engstler}\ \emph {et~al.}(1988)\citenamefont
  {Engstler}, \citenamefont {Krauss}, \citenamefont {Neldner}, \citenamefont
  {Rolfs}, \citenamefont {Schr{\" o}der},\ and\ \citenamefont
  {Langanke}}]{Eng88}%
  \BibitemOpen
  \bibfield  {author} {\bibinfo {author} {\bibfnamefont {S.}~\bibnamefont
  {Engstler}}, \bibinfo {author} {\bibfnamefont {A.}~\bibnamefont {Krauss}},
  \bibinfo {author} {\bibfnamefont {K.}~\bibnamefont {Neldner}}, \bibinfo
  {author} {\bibfnamefont {C.}~\bibnamefont {Rolfs}}, \bibinfo {author}
  {\bibfnamefont {U.}~\bibnamefont {Schr{\" o}der}},\ and\ \bibinfo {author}
  {\bibfnamefont {K.}~\bibnamefont {Langanke}},\ }\bibfield  {title} {\bibinfo
  {title} {Effects of electron screening on the ${}^3\mathrm{He}(d,
  p){}^4\mathrm{He}$ low-energy cross sections},\ }\href
  {https://doi.org/10.1016/0370-2693(88)90003-2} {\bibfield  {journal}
  {\bibinfo  {journal} {Physics Letters B}\ }\textbf {\bibinfo {volume}
  {202}},\ \bibinfo {pages} {179} (\bibinfo {year} {1988})}\BibitemShut
  {NoStop}%
\bibitem [{\citenamefont {Langanke}\ and\ \citenamefont {Rolfs}(1989)}]{Lan89}%
  \BibitemOpen
  \bibfield  {author} {\bibinfo {author} {\bibfnamefont {K.}~\bibnamefont
  {Langanke}}\ and\ \bibinfo {author} {\bibfnamefont {C.}~\bibnamefont
  {Rolfs}},\ }\bibfield  {title} {\bibinfo {title} {On the $t(d,n)\alpha$
  reactivity in fusion reactors},\ }\href
  {https://doi.org/10.1142/S0217732389002367} {\bibfield  {journal} {\bibinfo
  {journal} {Modern Physics Letters A}\ }\textbf {\bibinfo {volume} {4}},\
  \bibinfo {pages} {2101} (\bibinfo {year} {1989})}\BibitemShut {NoStop}%
\bibitem [{\citenamefont {de~Souza}\ \emph {et~al.}(2020)\citenamefont
  {de~Souza}, \citenamefont {Kiat}, \citenamefont {Coc},\ and\ \citenamefont
  {Iliadis}}]{de_Souza_2020}%
  \BibitemOpen
  \bibfield  {author} {\bibinfo {author} {\bibfnamefont {R.~S.}\ \bibnamefont
  {de~Souza}}, \bibinfo {author} {\bibfnamefont {T.~H.}\ \bibnamefont {Kiat}},
  \bibinfo {author} {\bibfnamefont {A.}~\bibnamefont {Coc}},\ and\ \bibinfo
  {author} {\bibfnamefont {C.}~\bibnamefont {Iliadis}},\ }\bibfield  {title}
  {\bibinfo {title} {Hierarchical bayesian thermonuclear rate for the
  7be(n,p)7li big bang nucleosynthesis reaction},\ }\href
  {https://doi.org/10.3847/1538-4357/ab88aa} {\bibfield  {journal} {\bibinfo
  {journal} {The Astrophysical Journal}\ }\textbf {\bibinfo {volume} {894}},\
  \bibinfo {pages} {134} (\bibinfo {year} {2020})}\BibitemShut {NoStop}%
\bibitem [{\citenamefont {Foreman-Mackey}\ \emph {et~al.}(2013)\citenamefont
  {Foreman-Mackey}, \citenamefont {Hogg}, \citenamefont {Lang},\ and\
  \citenamefont {Goodman}}]{Foreman_Mackey_2013}%
  \BibitemOpen
  \bibfield  {author} {\bibinfo {author} {\bibfnamefont {D.}~\bibnamefont
  {Foreman-Mackey}}, \bibinfo {author} {\bibfnamefont {D.~W.}\ \bibnamefont
  {Hogg}}, \bibinfo {author} {\bibfnamefont {D.}~\bibnamefont {Lang}},\ and\
  \bibinfo {author} {\bibfnamefont {J.}~\bibnamefont {Goodman}},\ }\bibfield
  {title} {\bibinfo {title} {emcee: {T}he {MCMC} {H}ammer},\ }\href
  {https://doi.org/10.1086/670067} {\bibfield  {journal} {\bibinfo  {journal}
  {Publications of the Astronomical Society of the Pacific}\ }\textbf {\bibinfo
  {volume} {125}},\ \bibinfo {pages} {306–312} (\bibinfo {year}
  {2013})}\BibitemShut {NoStop}%
\bibitem [{\citenamefont {Goodman}\ and\ \citenamefont
  {Weare}(2010)}]{goodman2010}%
  \BibitemOpen
  \bibfield  {author} {\bibinfo {author} {\bibfnamefont {J.}~\bibnamefont
  {Goodman}}\ and\ \bibinfo {author} {\bibfnamefont {J.}~\bibnamefont
  {Weare}},\ }\bibfield  {title} {\bibinfo {title} {Ensemble samplers with
  affine invariance},\ }\href {https://doi.org/10.2140/camcos.2010.5.65}
  {\bibfield  {journal} {\bibinfo  {journal} {Commun. Appl. Math. Comput.
  Sci.}\ }\textbf {\bibinfo {volume} {5}},\ \bibinfo {pages} {65} (\bibinfo
  {year} {2010})}\BibitemShut {NoStop}%
\bibitem [{\citenamefont {Barker}(1997)}]{Barker97}%
  \BibitemOpen
  \bibfield  {author} {\bibinfo {author} {\bibfnamefont {F.~C.}\ \bibnamefont
  {Barker}},\ }\bibfield  {title} {\bibinfo {title} {${\frac{3}{2}}^{+}$ levels
  of ${}^{5}${H}e and ${}^{5}${L}i, and shadow poles},\ }\href
  {https://doi.org/10.1103/PhysRevC.56.2646} {\bibfield  {journal} {\bibinfo
  {journal} {Phys. Rev. C}\ }\textbf {\bibinfo {volume} {56}},\ \bibinfo
  {pages} {2646} (\bibinfo {year} {1997})}\BibitemShut {NoStop}%
\bibitem [{\citenamefont {Azuma}\ \emph {et~al.}(2010)\citenamefont {Azuma},
  \citenamefont {Uberseder}, \citenamefont {Simpson}, \citenamefont {Brune},
  \citenamefont {Costantini}, \citenamefont {de~Boer}, \citenamefont
  {G\"orres}, \citenamefont {Heil}, \citenamefont {LeBlanc}, \citenamefont
  {Ugalde},\ and\ \citenamefont {Wiescher}}]{Azuma2010}%
  \BibitemOpen
  \bibfield  {author} {\bibinfo {author} {\bibfnamefont {R.~E.}\ \bibnamefont
  {Azuma}}, \bibinfo {author} {\bibfnamefont {E.}~\bibnamefont {Uberseder}},
  \bibinfo {author} {\bibfnamefont {E.~C.}\ \bibnamefont {Simpson}}, \bibinfo
  {author} {\bibfnamefont {C.~R.}\ \bibnamefont {Brune}}, \bibinfo {author}
  {\bibfnamefont {H.}~\bibnamefont {Costantini}}, \bibinfo {author}
  {\bibfnamefont {R.~J.}\ \bibnamefont {de~Boer}}, \bibinfo {author}
  {\bibfnamefont {J.}~\bibnamefont {G\"orres}}, \bibinfo {author}
  {\bibfnamefont {M.}~\bibnamefont {Heil}}, \bibinfo {author} {\bibfnamefont
  {P.~J.}\ \bibnamefont {LeBlanc}}, \bibinfo {author} {\bibfnamefont
  {C.}~\bibnamefont {Ugalde}},\ and\ \bibinfo {author} {\bibfnamefont
  {M.}~\bibnamefont {Wiescher}},\ }\bibfield  {title} {\bibinfo {title} {Azure:
  An $r$-matrix code for nuclear astrophysics},\ }\href
  {https://doi.org/10.1103/PhysRevC.81.045805} {\bibfield  {journal} {\bibinfo
  {journal} {Phys. Rev. C}\ }\textbf {\bibinfo {volume} {81}},\ \bibinfo
  {pages} {045805} (\bibinfo {year} {2010})}\BibitemShut {NoStop}%
\bibitem [{\citenamefont {Vousden}\ \emph {et~al.}(2015)\citenamefont
  {Vousden}, \citenamefont {Farr},\ and\ \citenamefont
  {Mandel}}]{Vousden_2015}%
  \BibitemOpen
  \bibfield  {author} {\bibinfo {author} {\bibfnamefont {W.~D.}\ \bibnamefont
  {Vousden}}, \bibinfo {author} {\bibfnamefont {W.~M.}\ \bibnamefont {Farr}},\
  and\ \bibinfo {author} {\bibfnamefont {I.}~\bibnamefont {Mandel}},\
  }\bibfield  {title} {\bibinfo {title} {Dynamic temperature selection for
  parallel tempering in {M}arkov chain {M}onte {C}arlo simulations},\ }\href
  {https://doi.org/10.1093/mnras/stv2422} {\bibfield  {journal} {\bibinfo
  {journal} {Monthly Notices of the Royal Astronomical Society}\ }\textbf
  {\bibinfo {volume} {455}},\ \bibinfo {pages} {1919–1937} (\bibinfo {year}
  {2015})}\BibitemShut {NoStop}%
\end{thebibliography}%
\end{document}